\documentclass[twocolumn]{article}

\usepackage{ae,aecompl}

\usepackage[english]{babel}
\usepackage[T1]{fontenc}
\usepackage{newtxtext,newtxmath}
\usepackage{booktabs,caption}
\usepackage[flushleft]{threeparttable}
\usepackage[a4paper,top=3cm,bottom=2cm,left=1.0cm,right=1.0cm,marginparwidth=1.0cm]{geometry}

\usepackage{amsmath}
\usepackage{graphicx}
\usepackage[colorinlistoftodos]{todonotes}
\usepackage{hyperref}
\usepackage{cleveref}
\usepackage{amssymb}	
\usepackage{multirow}
\usepackage[left]{lineno}
\usepackage{natbib}

\newcommand{\miniG}{mini-GWAC}
\newcommand{\GWAC}{GWAC}
\title{The mini-GWAC optical follow-up of the gravitational wave alerts:\\ results from the O2 campaign and prospects for the upcoming O3 run.
}

\author{D. Turpin$^{1}$$^\dagger$,
	C. Wu$^{1}$,
	X. H. Han$^{1}$,
	L. P. Xin$^{1}$,
	S. Antier$^{2,3}$,
	N. Leroy$^{2}$,
	L. Cao$^{1}$,
	H. B. Cai$^{1}$,
	B. Cordier$^{4}$,\and
	J. S. Deng$^{1,5}$,
	W. L. Dong$^{1}$,
	Q. C. Feng$^{1}$,
	L. Huang$^{1}$,
	L. Jia$^{1}$,
	A. Klotz$^{6,7}$,
	C. Lachaud$^{3}$,
	H. L. Li$^{1}$,\and
	E. W. Liang$^{8}$,
	S. F. Liu$^{1}$,
	X. M. Lu$^{1}$,
	X. M. Meng$^{1}$,
	Y. L. Qiu$^{1}$,
	H. J. Wang$^{1}$,
	J. Wang$^{8,1}$,
	S. Wang$^{1}$,\and
	X. G. Wang$^{8}$,
	J. Y. Wei$^{1,5}$,
	B. B. Wu$^{9}$,
	Y. J. Xiao$^{1}$,
	D. W. Xu$^{1,5}$,
	Y. Xu$^{1}$,
	Y. G. Yang$^{10}$,
	P. P. Zhang$^{1}$,\and
	R. S. Zhang$^{1}$,
	S. N. Zhang$^{1}$,
	Y. T. Zheng$^{1}$,
	S. C. Zou$^{1}$
	\\
}
\date{}
\begin{document}
	\twocolumn[
	\begin{@twocolumnfalse}
	\maketitle

	$^{1}$\normalsize{Key Laboratory of Space Astronomy and Technology, National Astronomical Observatories, Chinese Academy of Sciences, Beijing 100101}\\
	$^{2}$\normalsize{LAL, Univ. Paris-Sud, CNRS/IN2P3, Universit\'e Paris-Saclay, F-91898 Orsay, France}\\
	$^{3}$\normalsize{APC, Univ Paris Diderot, CNRS/IN2P3, CEA/lrfu, Obs de Paris, Sorbonne Paris Cit\'e, France}\\
	$^{4}$\normalsize{CEA Saclay, DRF/IRFU/D\'epartement d'astrophysique, 91191 Gif-sur-Yvette, France}\\
	$^{5}$\normalsize{School of Astronomy and Space Science, University of Chinese Academy of Sciences, 101408 Beijing}\\
	$^{6}$\normalsize{Universit\'e de Toulouse, IRAP 14 Av. Edouard Belin, F-31000 Toulouse France}\\
	$^{7}$\normalsize{Institut de Recherche en Astrophysique et Plan\'etologie (IRAP), UPS-OMP, Toulouse, France}\\
	$^{8}$\normalsize{Department of Physics and GXU-NAOC Center for Astrophysics and Space Sciences, Guangxi University, Nanning 530004, China}\\
	$^{9}$\normalsize{Institute of High Energy Physics/CAS, 19B YuquanLu, Beijing, 100049 , China}\\
	$^{10}$\normalsize{School of Physics and Electronic Information, Huaibei Normal University, Huaibei
		235000, China}
	\\
$^\dagger$ dturpin@nao.cas.cn
\begin{center}
accepted on Aug. 2019 in Research in Astronomy and Astrophysics
\end{center}
	
\begin{abstract}
	The second (O2) observational campaign of gravitational waves (GW) organized by the LIGO/Virgo Collaborations \textcolor{black}{has led} to several breakthroughs \textcolor{black}{such as} the detection of gravitational {wave} signals from {merger} systems involving black holes or neutrons stars. During O2, 14 gravitational {wave} alerts were sent to the astronomical community with sky regions covering mostly over hundreds of {square} degrees. Among them, 6 have been \textcolor{black}{finally} confirmed as real {astrophysical} events.
	Since 2013, a new set of ground-based {robotic} telescopes called \GWAC{} ({Ground Wide field Angle Cameras}) and its pathfinder \miniG{} \textcolor{black}{have been developed} to contribute to the various challenges of the multi-messenger and time domain astronomy. The GWAC system is \textcolor{black}{built up in} the framework of the ground-segment system of the SVOM mission {that will be devoted} to the study \textcolor{black}{of} the multi-wavelength transient sky in the next decade. During O2, \textcolor{black}{only} the \miniG{} telescope network was fully operational. Due to \textcolor{black}{the} wide field of view and fast automatic follow-up {capabilities} \textcolor{black}{of} the \miniG{} telescopes, \textcolor{black}{they} were well adapted to efficiently cover the sky localization \textcolor{black}{areas} of the gravitational wave event candidates. In this paper, we present the \miniG{} pipeline we have set up to {respond} to the GW alerts and we report our optical \textcolor{black}{follow-up} observations of 8 \textcolor{black}{GW alerts} detected during the O2 run. Our observations provided the largest coverage of the GW localization {areas} in a short latency made by any optical {facility}. We found tens of optical transient candidates in our images, but none of those could be securely {associated with} any confirmed black hole - black hole merger event. Based on this first experience and the near future technical improvements of our network system, we will be more competitive to detect the optical counterparts from some gravitational wave events \textcolor{black}{that will be} detected during the upcoming O3 run, especially those \textcolor{black}{emerging} from binary neutron star mergers.
\end{abstract}


\end{@twocolumnfalse}
]
\section{Introduction}           
\label{sec:intro}

The new generation of gravitational wave (GW) LIGO/Virgo detectors have given us an access to a new physics on the compact and extreme objects in the Universe such as the black holes (BH) or the neutron stars (NS) with unprecedented details, see for example \citep{LIGO16a}.
In 2015, the O1 \textcolor{black}{GW} observational campaign, marked the \textcolor{black}{birth} of the gravitational {wave} astronomy with the first two detections of GW signals {produced by} the coalescence of black holes bounded in \textcolor{black}{binary systems} (BBH) \citep{LIGO16b,LIGO16c}. A search for electromagnetic counterparts from these merger systems was \textcolor{black}{performed} without any significant result. While {any} electromagnetic {counterpart} from \textcolor{black}{a} BBH merger \textcolor{black}{event is} very unlikely, it has not been completely ruled out by some models under particular conditions \citep{Loeb16,Zhang16, Zhang16b,Perna16,Mink17}. In addition to that, the poor \textcolor{black}{localization} of these GW events and the long delay \textcolor{black}{of} the alert communication dramatically reduced the detection capabilities of the electromagnetic facilities. From November 2016 to August 2017, the O2 run has been effective for almost one year with a release of 14 alerts to the external partners of the LIGO/Virgo Collaborations (LVC). This leads to new discoveries of gravitational waves from compact mergers \citep{LIGO19}. In particular, on 17$^{th}$ August 2017, the discovery of the GW signal GW170817 emitted, for the first time, from the inspiral and the merger of two neutrons stars (BNS) marked the dawn of the multi-messenger astronomy \citep{LIGO17a,LIGO17b, LIGO17c}. Two matter ejecta were identified after this merger. First, almost simultaneously to the GW signal, a short gamma-ray burst (sGRB), GRB170817A \citep{Goldstein17}, and much later its associated X-ray and radio afterglows as long as the relativistic ejecta heats up its surrounding environment \citep[for a review on sGRB see][and references therein]{Berger14}. \textcolor{black}{Secondly}, about 10 hours after the GW trigger time, thanks to the intensive follow-up observations made by various optical facilities, an isotrotropic ejecta was also clearly identified as the signature of r-processes occuring in a {so-called} kilonova ejecta as predicted years ago by several authors \citep[][for a recent review]{Li98,Kulkarni05,Metzger10,Metzger17}. GW170817 permits to validate for the first time the merger model {proposed} decades ago to explain the short gamma-ray burst phenomena \citep{Paczynski86,Eichler89,Paczynski91}.
Beyond this remarkable result, the O2 run demonstrated the importance of having a third detector with {the} advanced Virgo, entering in science mode, to significantly reduce the {error on the localization} of some GW events \citep{LIGO17a,LIGO17d}. 
\textcolor{black}{However,} the Virgo detector only joined the last month of the O2 run, thus, a large majority of the O2 GW candidates remained poorly localized. According to the online LVC detection pipeline, the median size of the sky localization error box of the O2 GW alerts was $\mathrm{\sigma_{90\%} =1725~deg^2}$ \citep{LIGO19}. Practically speaking, in the electromagnetic domain, with such \textcolor{black}{localization} constraint and depending on the distance of the event, the discovery potential of the telescopes \textcolor{black}{having} relatively small field of views (typically FoV $<$1 sq.deg.) \textcolor{black}{and usually operated} in pointing mode is very low. \textcolor{black}{As a consequence, } it was primordial to conduct efficient electromagnetic follow-up\textcolor{black}{s} \textcolor{black}{using optimized strategies for both small and wide field of view telescopes}. The electromagnetic counterpart searches were therefore performed through various observational strategies \textcolor{black}{including archival} data analysis, prompt searches with all-sky instruments, wide-field tiled searches, targeted searches of potential host galaxies with small field of view facilities, and deep follow-up of individual sources. In the optical domain, the wide field instruments have the advantage of being able to cover a large fraction of the GW error boxes in a minimum \textcolor{black}{amount} of time. \\
Since 2013, the Ground-based Wide field Angle Cameras (\GWAC{}) \textcolor{black}{telescopes} are under development at the Xinglong Observatory in China to prepare the future ground segment of the SVOM mission dedicated to the study of the transient sky in 2021 with \textcolor{black}{both spaced-based} and ground-based multi-wavelength instruments \citep{SVOM16}. Due to the design of its extreme wide field of view ($25^\circ\times 25^\circ$), \textcolor{black}{the} \GWAC{} \textcolor{black}{telescopes} are well suited for the optical follow-up of the GW candidates. They have the capability to perform routine observations of the transient sky every night and, as being robotic, they are able to cover very rapidly a significant portion of the GW {localization regions}. These two specificities \textcolor{black}{allowed} us to conduct the first extensive {optical} follow-up of gravitational wave events, {searching for early optical counterparts,} from China. For the O2 GW run, \textcolor{black}{our optical follow-up} campaign was performed with the pathfinder \textcolor{black}{telescopes} \miniG{}.
\smallbreak
In this paper we present our \textcolor{black}{optical follow-up system of} the O2 GW alerts and the results of our campaign. We will firstly describe, in section \ref{sec:mini-GWAC_system}, our \miniG{} telescopes used during O2. We then present, in section \ref{sec:transient_program}, our transient research program set up {to respond} to any multi-messenger alerts. The results of our follow-up observations of the gravitational wave alerts are shown in section \ref{sec:GWAC_followup}. In section \ref{sec:O3}, we will discuss the {improvements} of our {detection} capabilities for the upcoming O3 run. Finally, we draw our conclusion in section \ref{sec:Conclusion}.

\section{The \miniG{} telescopes}
\label{sec:mini-GWAC_system} 
\begin{figure*}[h!]
	\centering
	\includegraphics[scale = 0.07]{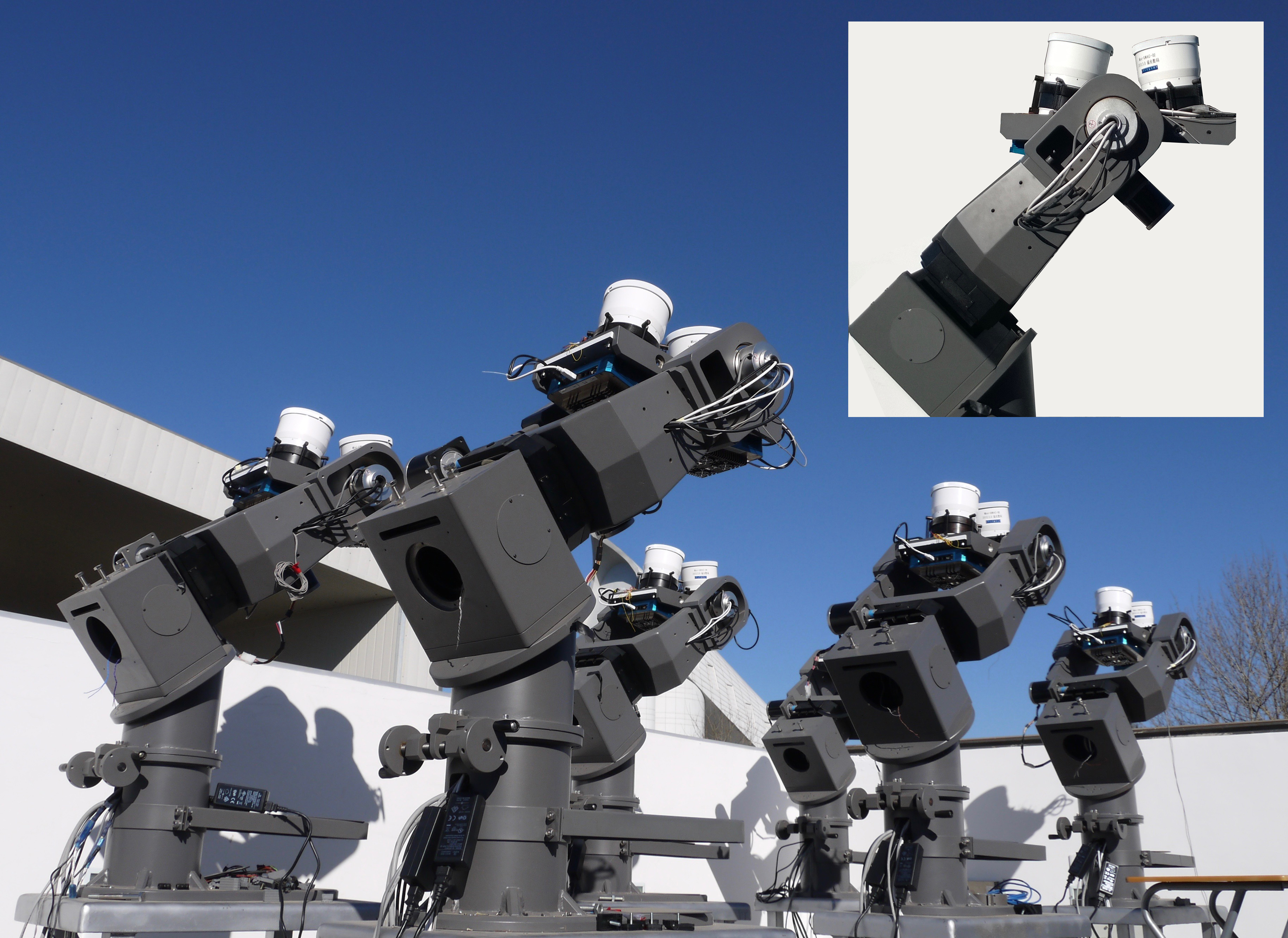}
	\includegraphics[scale = 0.40]{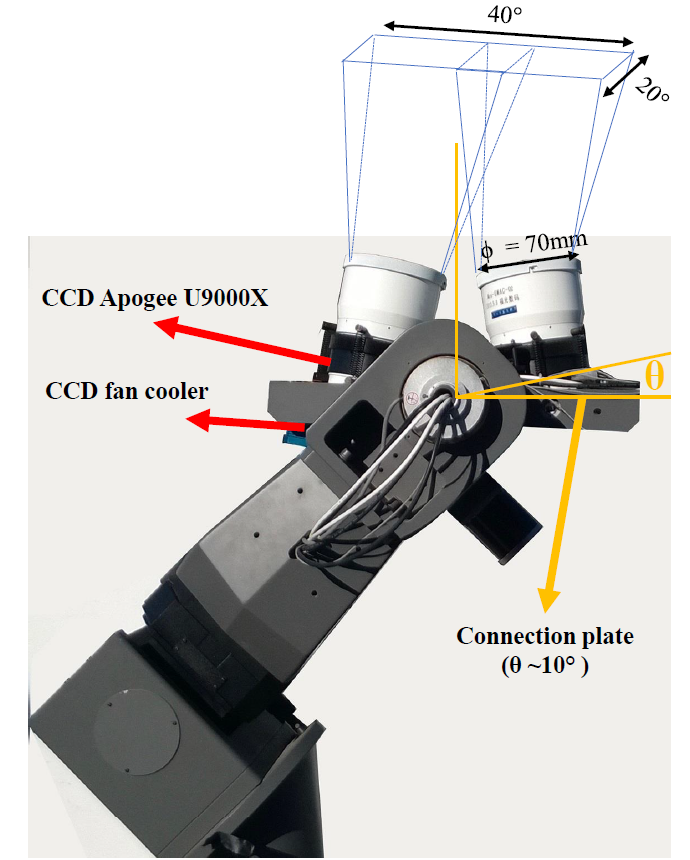}
	\caption{{(\textit{left}) The \miniG{} telescope farm, located at the Xinglong Observatory, includes 6 mounts and 12 Canon 85/f1.2 cameras. (\textit{Right}) Each mount is equipped with 2 cameras with a field of view (FoV) of 20$^\circ\times 40^\circ$ for a total FoV for the whole system of about 5000 sq.deg (about 1/4 of the Northern sky). The image cadency is 15 seconds.}}
	\label{Fig:minigwac}
\end{figure*}

In 2013, a \textcolor{black}{GWAC pathfinder}, called \miniG{}, has been developed in order to test and validate both the hardware and the data processing pipeline of the future GWAC system.\\
Located at the Xinglong Observatory (lat = 40$^\circ$23'39''N, lon = 117$^\circ$34'30''E) and founded by the National Astronomical Observatories (NAOC, Chinese Academy of Sciences), the \miniG{} network \textcolor{black}{is composed of} 6 mounts. Each mount is equipped with 2 Canon 85/f1.2 cameras {with an aperture of 7 cm}, as shown in Figure~\ref{Fig:minigwac}. 
For each camera, the detector is a CCD Apogee U9000X\footnote{{More details on the CCD detector can be found here: \url{http://www.lulin.ncu.edu.tw/slt40cm/U9000.pdf}.}} with an image cadence of 15 seconds (exposure=10s, read-out=5s) {and a readout noise of 12 electrons RMS at 1 MHz. Each camera is cooled down to -45$^\circ$ C with respect to the \textcolor{black}{local} environment temperature with a thermoelectric cooler system with forced air}. Two cameras are installed on a connection plate with a fix angle and are paved in a rectangle sky field. With such a configuration, one mount has a field of view of 20 degrees along the longitude direction and 40 degrees along the latitude one. This results in a field of view (FoV) of 800 square degrees per mount. Combining the network of the 6 mini-GWAC mounts, the overall FoV is about 5000 square degrees.  From the \miniG{} single images, a typical limiting (unfiltered) magnitude of about  12 is obtained in \textcolor{black}{a} dark night without clouds.
The \miniG{} {telescopes} have been designed with an extreme wide field of view and a small imaging cadence in order to mainly search for short-time scale optical transients {(OTs)}. The first light of \miniG\ was obtained on October 2015 during the O1 GW science run and the first follow-up of a GW event was made for GW151226 \citep{Wei15a}. A specific data processing pipeline has been developed to automatically detect in real-time {OT} candidates in the images. \\

\begin{figure}[h!]
	\centering
	\includegraphics[trim= 30 50 0 50, clip=true,scale = 0.6]{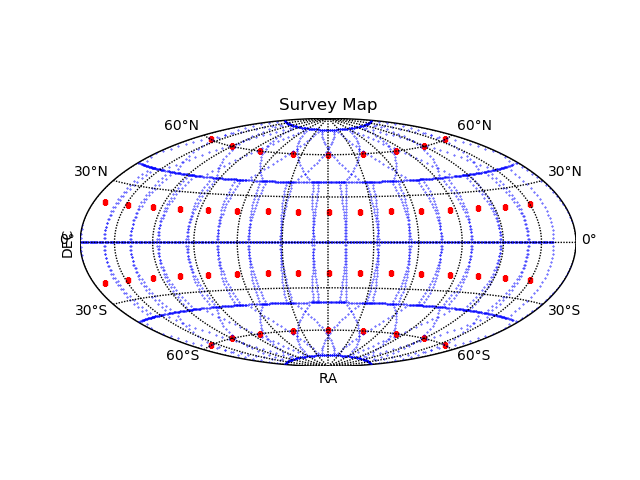}
	\caption{The sky, in Equatorial coordinates, fragmented in grids of equal area according to the \miniG\ per mount's field of view. Each night, observations are performed in a survey mode following the grid pointings (red dots) with a maximum of three grids per mount to be visited. At the position of the Xinglong observatory the grids with declination $\delta<20^\circ$ S are never observable. }
	\label{Fig:(mini)GWAC_grids}
\end{figure}

Each \miniG{} \textcolor{black}{telescope} is {operated} in a sky survey mode.
A pre-planed sky monitoring strategy is adopted, so that the \textcolor{black}{all} sky is partitioned into several fixed grids whose sizes are based on each mount's FoV, see Figure \ref{Fig:(mini)GWAC_grids}. 
During a night, each \textcolor{black}{telescope} {starts to monitor} one assigned sky grid until this one is no longer observable. {For a given mount, each observed grid is chosen to optimize its observational conditions, i.e. a high elevation above the horizon, a minimum distance to the moon of 20$^\circ$ when the moon phase is lower than 0.5 (half moon, 1 is the full moon phase) and 30$^\circ$ otherwise and also {having no overlap} with the other grid pointings \textcolor{black}{observed by other \miniG{} telescopes}. Once the first grids are no
	longer observable, the mounts {automatically slew} to {observe} new grids following the same observational strategy. Typically,} no more than three different grids are usually monitored by a single mount in a {single} night. During the observations, each camera is automatically focused to make the image quality at its best level following the method developed by \cite{Huang15}. The images taken by {all the} \miniG{} cameras are then analyzed in real-time and independently {camera per camera}.
\section{The \miniG{} optical transient search program}
\label{sec:transient_program}
During the \miniG{} {survey}, we simultaneously run a program dedicated to the discovery of new optical transient sources in our images. This search program relies on two main steps: the detection of the OT candidates and then their classification using various filters. The OTs that can be detected in our \miniG\ images originate from two classes of triggers: the external triggers such as \textcolor{black}{the} GW alerts or the {internal triggers, i.e the alerts produced by the GWAC system itself after the detection of an OT} in real-time by chance {in our images}. Typically, in the external \textcolor{black}{trigger} case, we expect to catch the early phases of the GRB afterglow emission, {some supernovae previously discovered by other groups}, galactic explosive events such as cataclysmic variables (CVs), tidal disruption events \textcolor{black}{or the optical counterparts from GW events}. For the {internal} triggers, we expect to rather detect near-Earth objects, uncatalogued flaring stars, supernovae, galactic transients and also many unexpected optical transients as the time-domain covered by \miniG{}/\GWAC{} (less than the minute timescale) is still largely unexplored yet in the optical domain. \\
The analysis of the images is performed in {real-time} \textcolor{black}{using} two transient search methods, i.e. the catalog cross-matching method and the difference \textcolor{black}{image} analysis \textcolor{black}{(DIA)}. These \textcolor{black}{methods} usually {yield} the detection of dozens of optical transient candidates by each \miniG{} telescope {every} night. \textcolor{black}{In the following section}, we briefly describe our two detection pipelines.\\


\subsection{The online mini-GWAC data processing}
\label{sec:data_proc}
\subsubsection{The catalog cross-matching method}
A specific pipeline to detect short-living transients in the \miniG{} images has been developed mainly from the IRAF\footnote{IRAF is distributed by NOAO, which is operated by AURA, Inc., under cooperative agreement with NSF.} package and SourceExtractor \textcolor{black}{software} \citep{Bertin96}.
The method is based on the comparison of the transient candidate positions found in the images with those of objects already catalogued in public archives. The catalog used in our pipeline is a mixture of the USNO B1.0 catalog and the stellar catalog produced \textcolor{black}{by SourceExtractor using} our reference images. The USNO B1.0 catalog has been chosen because of its all-sky coverage with reasonable astrometric measurements and a high completeness down to V=16, corresponding to the nominal design for the GWAC sensitivity. The reference images are obtained by co-adding 10 images of high quality from the same grid region. These images are automatically picked-up in the \miniG{} image database and selected based on the quality of their stellar point spread function (PSF), background brightness and atmospheric transparency.\\
{Note that the coma is quite serious at the extreme edge of the mini-GWAC images which affect our detection
	efficiency. We estimated a loss of about 0.5 mag \textcolor{black}{in our sensitivity threshold between OTs detected in} the
	extreme edge of the image, where the PSF of stars can slightly deviate from a 2D gaussian profile, and
	the inner part of it (typically the 2k x 2k part of the image).}
A new optical source is detected in our images if it fulfills the following criteria:
\begin{itemize}
\item[](i) The candidate must not be detected in the reference image with a signal-to-noise ratio greater than SNR = 5, while it is in the night images.
\item[](ii) In order to exclude some moving objects, the candidate shall be detected in at least two continuous images without any apparent shift in its position.
\item[](iii) There is no any minor planet object with a brightness larger than 13 mag near the location of the candidate. The choice of this limiting magnitude is made according to the sensitivity of the \miniG{} telescopes.
\item[](iv) There is no any defect in the CCD camera at the location of the candidate.
\item[](v) The PSF and the ellipticity of any candidate shall be stellar-like profile (2D gaussian profile {with a limited deviation). At the edge of the image, this criterion reduces our detection efficiency
	for faint sources.}

\end{itemize}

If {an} OT candidate is confirmed as being an {uncatalogued} source, then our pipeline {allows} to sample the optical emission of the transient in a short time resolution of 15 seconds. In order to improve our detection capabilities, a stacking analysis based on a \textcolor{black}{group} of ten images is also processed in parallel. This allows to increase the signal to noise ratio (SNR) of faint objects to detect them at the edge of our camera sensitivity but with a lower time resolution. For these faint OTs we will finally reach a time resolution from several minutes to {a} few hours.

\subsubsection{Differential image analysis}

The difference \textcolor{black}{image} analysis (DIA) is made \textcolor{black}{by} following three steps: 
\begin{itemize}
\item[](i) an image alignment between the reference and the night images.
\item[](ii) the difference between the two images to obtain a residual image.
\item[](iii) the transient candidate selection after the residual analysis.
\end{itemize}
First, for the image alignment method, we used the Becker implementation\footnote{\url{http://www.astro.washington.edu/users/becker/v2.0/hotpants.html}} of the \cite{Alard00} algorithm finely tuned for the \miniG{} data. All the images (reference and night) used for DIA are truncated from the $3056\times3056$px of the raw image to $2001\times2001$px to avoid the bad PSF quality near the edge of the images. Before the subtraction, \textcolor{black}{flux and PSF} calibrations are operated on both images to obtain the best residuals possible. Once the subtraction is made, the transient selection \textcolor{black}{program} employs a supervised machine learning routine based on a random forest algorithm to preliminary classify the spurious points in the residual images. The reference images are taken days before the trigger time to ensure, as much as possible, that no optical {precursor} is present in our data at the OT candidate position. Then, the OT selection criteria follow the same rules than the ones described above for the catalog cross-matching method. With such DIA method we can also apply a stacking analysis \textcolor{black}{in} the images \textcolor{black}{to enhance our optical flux sensitivity.}

\subsection{Optical transient classification}
\label{sec:OT_class}
\begin{figure*}[h!]
	\centering
	\includegraphics[trim = 0 100 0 40,clip=true, scale = 0.35]{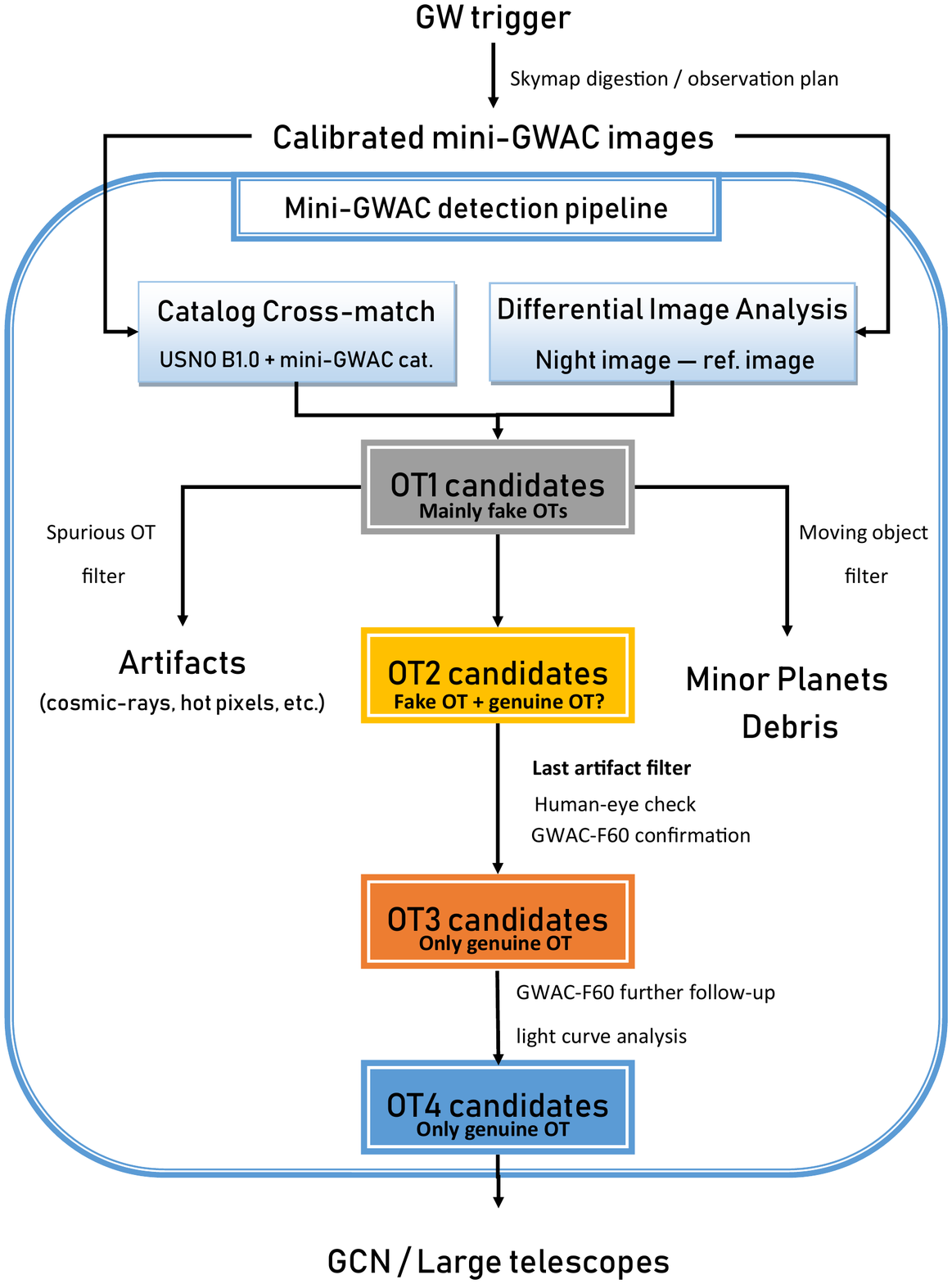}
	\caption{A schematic view of the \miniG{} detection pipeline for optical transients. Our pipeline identifies the optical transient candidates through different steps using both automatic and human actions.}
	\label{Fig:GWAC_pipeline}
\end{figure*}

Once an image is processed, a list of preliminary OT candidates is automatically established by comparing the subsequent results of the \textcolor{black}{two} detection pipelines. These candidates, labeled as \textit{OT1 candidates}, are \textcolor{black}{usually composed of} non astrophysical sources, fake optical transients such as minor planets or variable stars and \textcolor{black}{a few amount of} possibly genuine optical transient sources either in a rising or a fading phase.\\
The search for OTs then implies to carefully filter the OT1 candidates sample out of all the fakes through several steps.
The first series of selection criteria mostly {rely} on the PSF analysis of the candidates, additional checks in other all sky catalogs such as 2MASS, SDSS9, DSS2, and their detection in a time \textcolor{black}{series} of at least 2 images. From this step, most of the \textit{OT1 candidates} are \textcolor{black}{mainly} classified as non-astrophysical sources ( i.e hot pixel, crosstalk, cosmic-rays, dust and CCD artifacts, moving debris etc.) or astrophysical sources but identified as moving objects like minor planets. The candidates that pass these series of filters are then labeled as \textit{OT2 candidates}, the others are automatically rejected.\\
{The \textit{OT2 candidates} can still be a mix of fake OTs that were not well filtered during
	the first \textcolor{black}{steps} and few (or even zero) real OTs. {Therefore}, we analyze them} one by one through a human-eye check (PSF matching, lightcurve and public archive check). {For the candidates judged by our duty scientist as being promising, we trigger} fast extra multi-wavelength follow-up observations \citep[in prep.]{Yang19} {at deeper magnitudes (typically R $\sim$ 19 for an exposure = 120 seconds)} with two dedicated 60 cm robotic telescopes (GWAC-F60A/B, UBVRI filters, jointly operated by the NAOC and the Guangxi University). Based on this set of informations, {we may confirm some of the \textit{OT2 candidates} as being genuine optical transients, while the others are finally rejected. The remaining confirmed OTs are therefore labeled \textit{OT3 candidates}.} At this stage, we usually \textcolor{black}{reduce} the initial number of candidates per night and per \textcolor{black}{telescope} from dozens to {a very few (including zero)} for the \miniG{} system.\\
The \textit{OT3 candidates} are automatically {followed}-up as long as possible during the night to better characterize \textcolor{black}{the color evolution of their optical emission}. According to the evolution of their lightcurves, we may associate some of these OTs to the astrophysical event (a GW merger event for example) that had triggered such observations. If so, we will then publish an alert using the {Gamma-ray Coordinates Network \footnote{\url{https://gcn.gfsc.nasa.gov}} (GCN) system} and also quickly ask for spectroscopic follow-ups to the larger telescopes in China (2.16m at the Xinglong Observatory, 2.4m telescope at the Lijiang station of the Yunnan Observatory). Such very promising OT candidates constitute our final sample labeled \textit{OT4 candidates}.
Our detection pipeline is summarized in Figure \ref{Fig:GWAC_pipeline}.
After our selection process, the transient candidates are classified under six categories in our database: 
\begin{itemize}
	\item[]{\bf Category A / The sources already catalogued}: This category groups together the OT candidates that have finally matched \textcolor{black}{the positions of} known catalogued stars in the SIMBAD database \citep{Wenger00}. This database is complete for the limiting magnitude of the \miniG{} telescopes (V=12).\\
	\item[]{\bf Category B / The suspected variable/flaring stars}: These OT candidates are tagged as variable stars when their {positions} matched the one of {an} already catalogued variable star and their lightcurves evolution is in good agreement with the one of the \textcolor{black}{associated} variable star.\\
	\item[]{\bf Category C / The moving objects}: The candidates are identified as moving objects by their tracks in several images or if they are already catalogued in the Minor Planet data center\footnote{\url{https://minorplanetcenter.net//iau/mpc.html}}.\\
	\item[]{\bf Category D / The spurious points}:  This category groups together the OT candidates as being cosmic rays, instrument defects like hot pixels and noise in the residual images. The classification criteria are based on the occurrence rate of the source in our images. Typically, an OT candidate with an occurrence of less than twice in the image time series, its historical data and the residual image is identified as noise.\\
	
	\item[]{\bf Category E / The OTs with a host galaxy}: This category groups the \textit{OT3 candidates} that have matched, within a circle region of 90 arcsec around the \miniG{} position (corresponding to $\sim$3 \miniG{} pixels), the position of very nearby galaxies of the RC3 catalog \citep{Corwin94}. This catalog is complete enough at the \miniG{} limiting magnitude. This category actually may gather kilonovas (for the purpose of GW optical follow-up), supernovae, bright tidal disruption events, etc.\\
	
	\item[]{\bf Category F / The host-less OTs}: This category groups the \textit{OT3 candidates} having no match with the RC3 galaxy catalog. Typically, these candidates may correspond to host-less astrophysical events or extragalactic/cosmological events such as Gamma-ray Burst afterglows.
\end{itemize}

\subsection{{The detection efficiency of \miniG{} system}}

{The optical transient search program has run for several years from 2014 to 2017 (not continuously)
	and being updated every year. In this section, we aim to estimate the number optical transients \textcolor{black}{the \miniG{} telescopes are} able to serendipitously detect in single frames \textcolor{black}{according to} our archival data. Our analysis is based on the latest period of mini-GWAC operation when the
	detection pipeline was upgraded to its last version so that the perfomances could be compared to the
	period covered by the O2 run. We selected six months of data between Oct. 2016 and Mar. 2017
	which corresponds to a total amount of 1673607 images.}\\
{Within this period of archival data, 75 individual optical transient sources (typically flaring stars
	and few unclassified astrophysical optical transients) were detected by mini-GWAC in several hundreds
	of single frames. We therefore estimate that the expected number of new transient\textcolor{black}{s} per single
	frame is on average $\mathrm{N_{OT/frame} = 4.5 \times 10^{-5}~OT/frame}$. In other words, the mini-GWAC network
	is able to detect a new optical transient such as flaring stars brighter than $\rm{m_R} \sim$ 12 about
	every 11.5 days assuming that on average a night at Xinglong lasts 8 hours. For a single camera, one night corresponds to about
	1920 frames (including the readout time of 5 seconds for each frame). The OTs detected
	by one mini-GWAC camera can be considered as Poissonian events in our sky survey observations with a typical
	rate per night given by $\mathrm{\lambda = N_{OT/frame} \times 1920 ~OT/night}$. As a consequence, we estimate that
	the Poissonian probability of detecting at least one OT, brighter than $\rm{m_R} \sim$ 12, during a night with one camera is
	$\mathrm{P[N_{OT,night}\ge 1| \lambda= 8.6 \times 10^{-2}] \sim 8.2\%}$.\\
	A single frame catches a sky pattern of about 400 square degrees which finally gives the number
	of optical transient per square degree per frame exposure time one may expect to detect by chance with {one} \miniG{} {camera}:
	\begin{equation}
	\label{eq:eq1}
	\begin{split}
	\mathrm{N_{OT/sq.deg/\Delta T_{frame}} = \frac{N_{OT/frame}}{FoV_{RA} \times FoV_{dec}}= 1.1 \times 10^{-7}}\\
	\mathrm{~OT\cdot deg^{-2} \cdot \Delta T^{-1}_{frame}}
	\end{split}
	\end{equation}
	where $\mathrm{\Delta T_{frame}}$ = 10 seconds and $\mathrm{FoV_{RA} = FoV_{dec} = 20^\circ}$. We emphasize that these statistics
	have to be taken as rough estimates of the mini-GWAC perfomances since they are averaged on very different
	\textcolor{black}{observational} conditions (weather, sky brightness, moon distance, airmass, duration of the observations
	per night, etc.) and random source positions in the images for which the detection efficiency can vary
	between the edge and the inner part of the image, see \ref{sec:data_proc}. However, these statistics give the right order
	of magnitude and will be useful to understand the significance of any association of an OT detected in
	spatial coincidence with a gravitational wave event.
}

\section{The O2 follow-up campaign of \miniG{}}
\label{sec:GWAC_followup}

During the O2 GW observational campaign, 14 alerts have been sent to the external partners of the LIGO/Virgo Collaborations (LVC). The GW candidates were classified into two categories of potential astrophysicals events able to emit gravitational waves: \textcolor{black}{the} compact binary mergers including \textcolor{black}{black holes and/or neutron stars} on one hand, and the collapse of a massive star or magnetars instabilities \citep{Kotake06,Ott09,Gossan15,Mereghetti08} (mentionned as Burst) on the other hand.\\
The alerts with false alarm rates less than one per two months were distributed in the format of notices and circulars via private GCNs. The latency of the initial alert dissemination was ranging from 30 minutes to few hours due to the necessary human validation of the data quality. Regular updates of the localization error box of the candidates were sent by LIGO/Virgo few hours up to few months.
All the events were finally classified much later through an offline analysis performed by the LVC \citep{LIGO19}. All of the confirmed events originated from compact binary mergers and except GW170817, the only BNS merger, they were classified as BBH mergers.  

\subsection{Alert reception system with \miniG{}}
The GW alerts were received through the GCN system as described in \citep{LIGO19} and then \textcolor{black}{recomposed in a VOEvent format}. The GW \textcolor{black}{bayesian probability} skymaps were decomposed using the predefined \miniG{} sky grids. \textcolor{black}{A list of tiles were therefore scheduled for observations by order of priority based on their respective probability of containing the GW event}. The observation plan was performed for each telescope so that the different \textcolor{black}{tiles} can be observed several times during the night.\\
The recomposed alerts were produced by our french science center located at the Laboratoire  de l'Acc\'el\'erateur Lin\'eaire (LAL) institute in Paris-Orsay and transmitted to the NAOC at Beijing into the chinese science center that operates our telescopes at the Xinglong Observatory. The message transfer connection \textcolor{black}{was} built with our own scripts developed in python language based on pub/sub mode of zeroMQ, which has features of authentication, encryption, and validation {of the} messages. The connection \textcolor{black}{protocol} also supports automatic re-connection and re-sending message. The typical latency time is $\sim0.16$ s. \textcolor{black}{Taking into account the} additional \textcolor{black}{ delays due to the} parsing and \textcolor{black}{the rewriting of the VOEvent} alert {as well as the} response {delay} of the telescopes, the total latency \textcolor{black}{for the alert receipt by \miniG{} was} \textcolor{black}{typically} less than 2 minutes.
\subsection{Our observations with mini-GWAC}
During the O2 campaign, \textcolor{black}{the \miniG{} telescopes} followed-up 8/14 gravitational waves alerts as shown in Figure \ref{Fig:GW_skymaps}. The {localization regions} of the six other GW alerts were not visible at the Xinglong Observatory at all.

From our eight successful follow-ups, two of them (GW170104, GW170608) were confirmed as {GW sources originated from} the inspiral and the merger of two black holes. The six remaining events were later retracted \citep{LIGO19}. The main results of our observational campaign are summarized in Table \ref{tab:GWAC_obs}. 

\begin{table*}[h!]
	\caption{Summary of the observations made at the Xinglong observatory during the O2 GW run with the \miniG{} telescopes.}
	\label{tab:GWAC_obs}
	\scriptsize
	\begin{center}
	\begin{tabular}{llclclcccl}
		\hline\hline\noalign{\smallskip}
		\multicolumn{4}{c}{Gravitational wave triggers} & & \multicolumn{5}{c}{\miniG{} observations}\\\cline{1-4}
		\cline{6-10}
		ID & $~~~~~~~~~$Trigger date  & Loc. error & confirmed/type & & $~~~~~~~~\mathrm{T_{start}}$ & $\mathrm{\Delta T_{obs}}$  & $\mathrm{P_{GW,cov}}$ & N$_{\rm{OT2}}$ &GCN Reference\\
		&  $~~~~~~~~~~~~$(UTC)  &   (90\%) deg$^2$     &  &  &  &  (h) & on $\mathrm{\Delta T_{obs}}$  & (MP tag) & \\
		(1) & $~~~~~~~~~~~~~~~$(2)  & (3) & $~~~~~~~~$(4) &  & $~~~~~~~~~~~$(5) & (6)  & (7) & (8) &$~~~~~~~~$(9)\\
		\hline
		G268556$^{(1)}$ & 2017-01-04 10:11:58 & 1630 & Yes / BBH & & $\mathrm{T_{GW}+2.3~h}$ & $\sim$10.0 & 62.4$\%$ & 273 (2) &\citealt{Wei17a}\\ 
		G270580 & 2017-01-20 12:30:59.35 & 3120 & No / Burst & & $\mathrm{T_{GW}+20~ min}$ & $\sim$9.5 & 53.8$\%$ & 30 (1) &\citealt{Wei17b}\\
		G274296 & 2017-02-17 06:05:55.05 & 2140 & No / Burst & & $\mathrm{T_{GW}+6.3~h}$ & $\sim$5.0 & 63.8$\%$ & 5 (3) &\citealt{Wei17c}\\
		G275404 & 2017-02-25 18:30:21 & 2100 & No / NS-BH & & $\mathrm{T_{GW}-5.5~h}$ & $\sim$9.0 & 31.7$\%$ & 88 (3) &\citealt{Wei17d}\\
		G275697 & 2017-02-27 18:57:31 & 1820 & No / BNS & & $\mathrm{T_{GW}+2.7~d}$ & $\sim$7.0 & 6.4$\%$ & 0 &\citealt{Wei17e}\\
		G277583 & 2017-03-13 22:40:09.59 & 12140 & No / Burst & & $\mathrm{T_{GW}+12.5~h}$ & $\sim$10.0 & 46.2$\%$ & 198 (8) &\citealt{Wei17f}\\
		G284239 & 2017-05-02 22:26:07.91 & 3590 & No / Burst & & $\mathrm{T_{GW}+2.6~d}$ & $\sim$8.0 & 22.0$\%$ & 47 (0) & \citealt{Xin17}\\
		G288732$^{(2)}$ & 2017-06-08 02:01:16.492  & 860 & Yes / BBH & & $\mathrm{T_{GW}+15~h}$ & $\sim$2.5 & 18.5$\%$ & 8 (0) &\citealt{Leroy17} \\
		\noalign{\smallskip}\hline
	\end{tabular}
	\end{center}
	\textit{Note:} The latency of the first image with the GW trigger time takes into account the GW alert transmission delay by the LVC to the multi-messenger community as well as the delay due to our own system and the local weather conditions. (3) See \citep{LIGO19}. (6) is the duration of the \miniG{} observations related to each trigger. (7) is the bayesian probability (integrated over our observation time) that the GW source is in our images based on the final release of the GW Bayestar skymap. (8) is the number of optical transient candidates (OT2) found during $\mathrm{\Delta T_{obs}}$ in the GW sky localization area (90\% C.L.). {None of these candidates were finally classified as real OT and so be credibly related to any GW event.} The numbers of OT candidates identified as minor planets are indicated in parenthesis. $^{(1)}$ renamed GW170104; $^{(2)}$ renamed GW170608.
\end{table*}

\subsubsection{Response latencies to the O2 GW alerts}
\begin{figure}[h!]
	\centering
	\includegraphics[width=0.8\columnwidth]{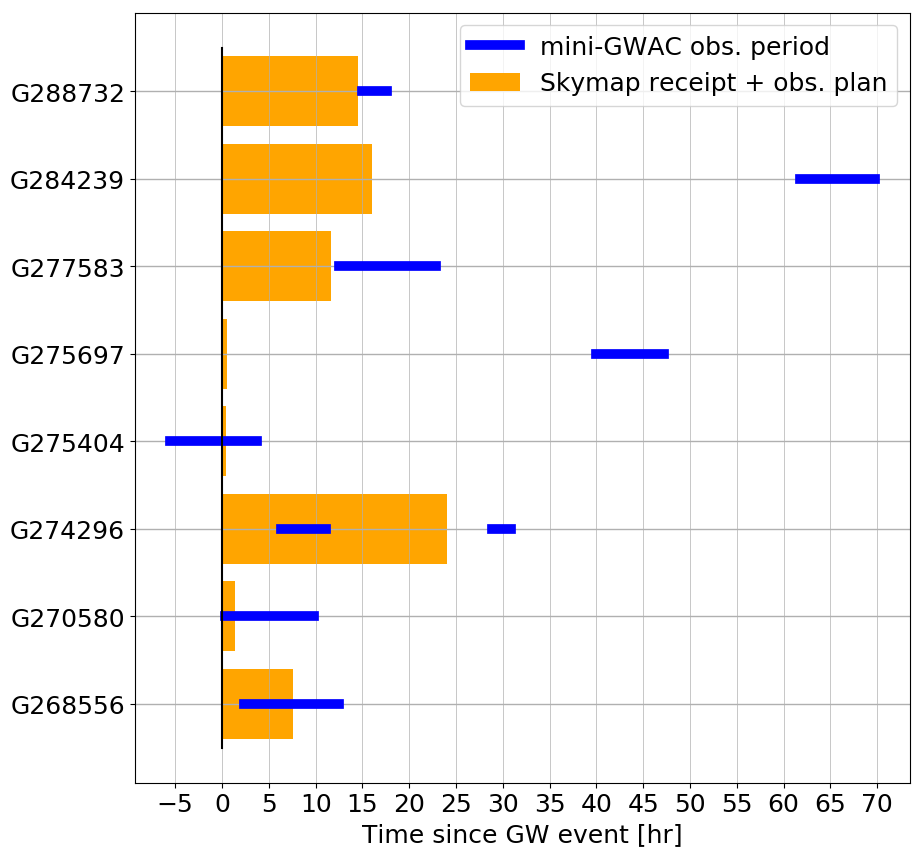}
	\caption{The \miniG{} response latencies to the GW alerts. For each GW event followed-up by \miniG{} during O2, the orange bars correspond to the delivery time of the alert at the Xinglong Observatory. This delivery time is mainly due to the time for the LIGO/Virgo Collaboration to send the circular alerts plus the time for our alert system to digest the GW skymap and produce an observation plan for \miniG{}. The blue bars show the period of our observations with respect to the GW trigger time.}
	\label{Fig:delays}
\end{figure}

Except for two events (G275697 and G284239) where the weather conditions prevented us from observing as soon as the GW trigger was received, we \textcolor{black}{responded} with a short latency to the GW alerts, typically within few minutes after the receipt of the alert \textcolor{black}{messages}. We then continuously monitored the sky localization areas during several hours in the first night following the GW trigger times. For half of the followed-up GW alerts (G268556, G270580, G274296 and G275404), we were actually already observing a part of their sky localization areas during our survey program prior to the alert receipt (and even before the GW event for G275404), see Figure \ref{Fig:delays}. 
This highlights two major advantages of such wide {field of view} telescopes {observing} in survey mode. First, for a significant amount of alerts, they can make simultaneous (even prior for possible precusors) observations based on their regular observational schedule. This also prevents from having no prompt image in case of a failure of the alert receiver system. During our O2 campaign, we experienced two failures of our alert receiver system. For G274296, it had no impact on our follow-up as our \miniG{} telescopes were actually already monitoring a sky area that covered the {full} GW error box visible at the Xinglong Observatory. However, for G277583, we underwent an additional delay due to an internet connection loss to start our observations. Once the connection came back, we immediately pointed our \miniG{} mounts {to} the GW {sky} \textcolor{black}{regions}.\\ In a second hand, some images can usually be taken few hours and even days before the GW events in the survey mode, when no electromagnetic counterpart is much expected. Therefore, the wide field {of view} telescopes have {a} considerable {amount of} reference images available for a large fraction of the sky \textcolor{black}{which offers} the possibility to make a quick vetting or confirmation of the optical transient candidates that may be found after \textcolor{black}{some merger events} by several other facilities.\\

\subsubsection{Coverage of the GW sky localization area}
\begin{figure}[h!]
	\centering
	\includegraphics[width=0.8\columnwidth]{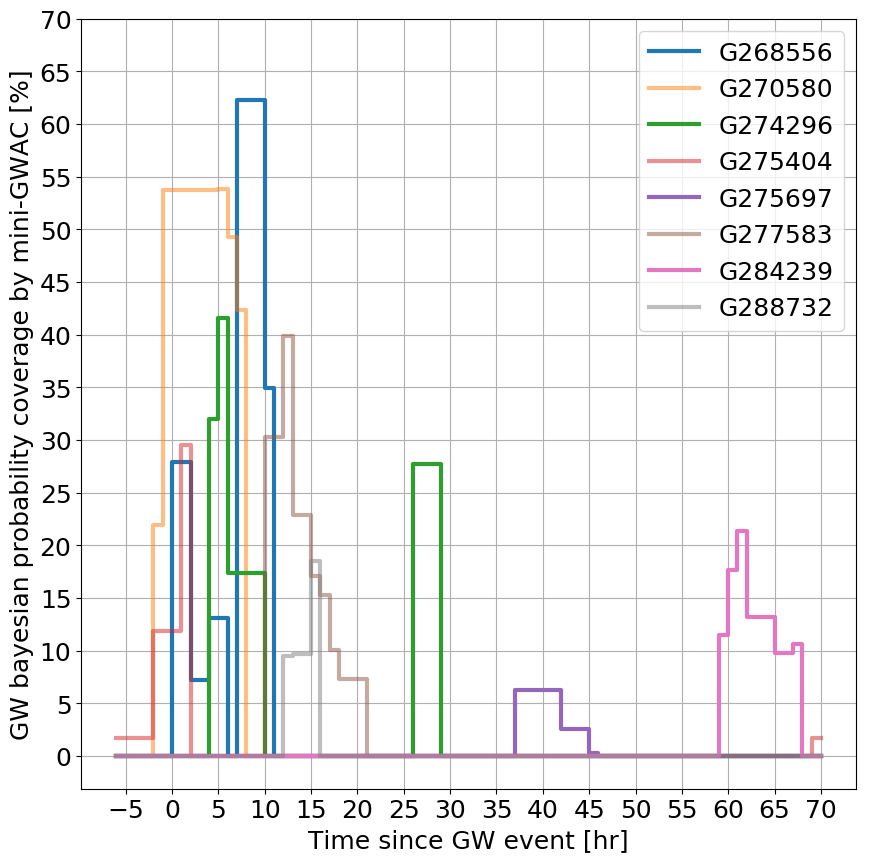}
	
	\caption{The evolution of our eight GW skymap coverages (bayesian probability) with \miniG{} as function of time expressed with the delay since the GW trigger time.}
	\label{Fig:GW_timecoverage}
\end{figure}

From the GW \textcolor{black}{bayesian probability} skymaps, we estimate that the median probability of having the GW events in our images during our periods of observation is 38.9$\%$. For some events, mainly located in the Northern hemisphere, our observations covered more than 60$\%$ of the bayesian localization. This is the largest coverage (based on a GW localization of several thousands of square degrees) performed by any optical telescope on a single night \textcolor{black}{during the O2 campaign}. We also computed the real-time performance of our follow-up system \textcolor{black}{concerning the} coverage of the bayesian probability skymaps as shown in Figure \ref{Fig:GW_timecoverage}.
During O2, our median instantaneous (based on periods of 1h of observation) bayesian probability coverage of the initial GW alert skymaps was $P_{cov,med} = 14.2 \%$. This quantity is much more representative of the real capabilities of our \miniG{} instruments to cover the GW localization area provided by only two interferometers (LIGO Handford and Livingston here). \textcolor{black}{It shows that despite} the active participation of the wide field of view telescopes to the follow-up campaign, such as the \miniG{} telescopes, the need to reduce the GW sky localization area is still crucial to optimise the scientific returns.

\subsection{Results}
\begin{figure}[h!]
	\centering
	\includegraphics[width=0.8\columnwidth]{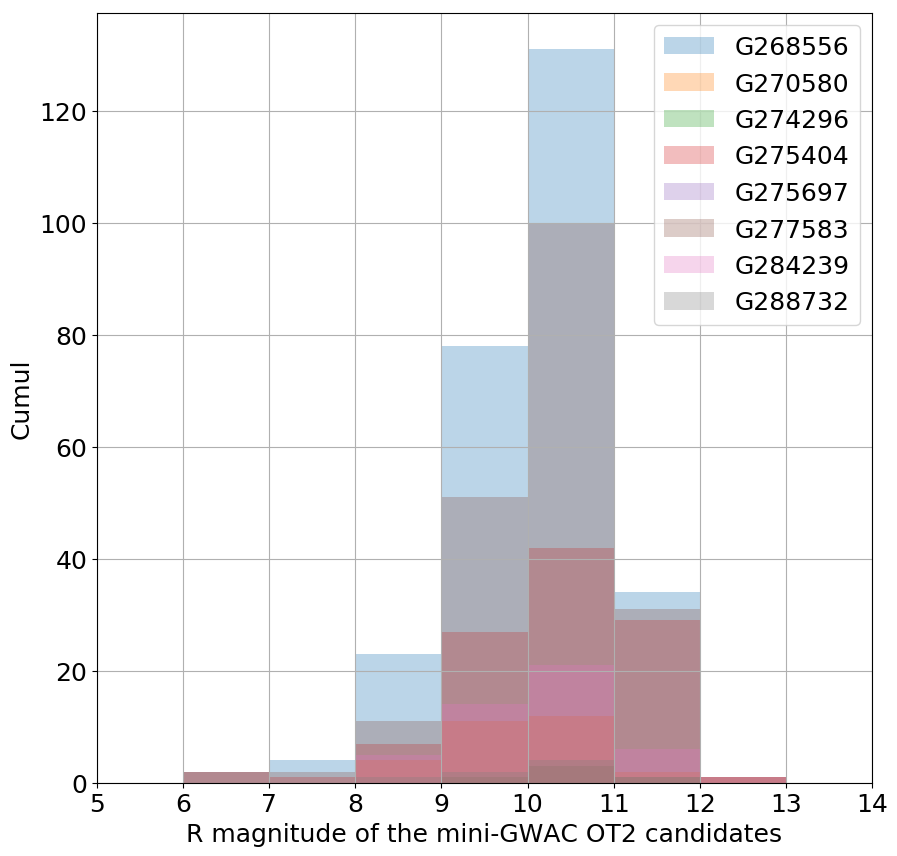}
	\caption{Distribution of the R-band magnitude (unfiltered calibrated with the USNO B1.0 R2mag catalog) of the optical transient (OT2) candidates found in \miniG{} images for each GW event. These magnitudes are computed at the time of the detection of the OT candidates.}
	\label{fig:mag_dist_OT}
\end{figure}

\begin{figure}[h!]
	\centering
	\includegraphics[width=0.3\columnwidth]{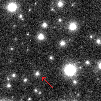}
	\includegraphics[width=0.3\columnwidth]{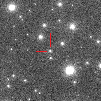}
	\includegraphics[width=0.3\columnwidth]{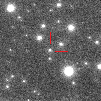}
	\caption{The reference (left) and the first and last night images (middle and right) of a moving object detected by \miniG{} during our follow-up of G274296 on 2017-02-17 12:18:28 (V = 11.1). Note that this minor planet (471 Papagena) is also present in the reference image (red arrow) a day before with an angular distance of about 13 arcmin with respect to its position measured during our observations.}
	\label{fig:OT_example}	
\end{figure}

The number of transient candidates found in our images and spatially correlated with the GW events depends on several parameters such as the size of the GW error box and our subsequent coverage of it, the duration of the observations \textcolor{black}{of} each grid as well as the local weather conditions (moon brightness, sky transparency, weather status, etc.). Taking {these factors} into consideration, we ended with more than 200 hundreds OT2 candidates for G268556 while, for example, we could not detect any credible transient source in our follow-up of G275697 (having the poorest coverage of all the GW events of our sample). In Appendix \ref{Appendix}, we give the details of our observations, grids per grids for each GW event. Our OT2 candidates are detected within a wide range of unfiltered magnitudes (calibrated in R-band Johnson Vega system) $m_R\in [12.3 - 6.8]$ see Figure \ref{fig:mag_dist_OT}.  
Concerning the two confirmed BBH merger events, GW170104 and GW170608, none of our \textit{OT2 candidates} (273 and 8, respectively). {were classified as real {OTs} and hence, no OT3 candidates emerged from this step. All of our OT2 candidates were finally {classified in the} category A (catalogued stars), category C (Minor planets) as shown in Figure \ref{fig:OT_example} or category D (spurious points). As a consequence we could unambiguously reject any association with the two merger events.} These null results can be explained both by observational constraints (sensitivity of our telescope, partial coverage of the GW error boxes) and by the physics of the BBH mergers that, if {they truly radiate an electromagnetic emission}, may power too faint optical transient emissions to be detected by our set of telescopes.
\\
We compared these null results with the number of optical transients we expected to find spatially correlated with the GW \textcolor{black}{skymaps} by chance in our period of observations. To do so, we used the following expression:
	\begin{equation}
	\begin{split}
	\mathrm{N^{serendipitous}_{OT,GW} = N_{OT/sq.deg/\Delta T_{frame}} \times f_{GW} \cdot \sigma^{90\%}_{GW} \times N_{frame}}
	\end{split}
	\end{equation}
	where $\mathrm{N_{OT/sq.deg/\Delta T_{frame}}}$ has been defined in equation \ref{eq:eq1}, $\rm{f_{GW}}$ is the fraction of the \textcolor{black}{GW} skymap we
	covered by our observations, $\mathrm{\sigma^{90\%}_{GW}}$ is the contour of the GW probability skymap given at the 90\%
	confidence level and $\mathrm{N_{frame}}$ is the number of single frames we took during our periods of observation.
	For each GW event, we actually computed this expression for every single tile covering a portion
	of the skymap during a certain amount of time, see our observation log in  Appendix \ref{Appendix}. For a given GW event, the final result is the addition of the expectations \textcolor{black}{given} in all the individual tiles for those that predict at least one event. \textcolor{black}{Otherwise}, if none of \textcolor{black}{the tiles} predict any OT detection, we took the best expectation among all the tiles. Concerning our observational campaign of the two BBH merger GW170104 and GW170608 we finally end up with $\mathrm{N^{serendipitous}_{OT,GW} \sim 2.6 \times 10^{-2}}$ and $\mathrm{6.0 \times 10^{-3}}$ expected OT, respectively. These estimates highlight the fact that any single OT detected in spatial coincident with any of these two GW events would have been
	of very great interest as a serendipitous \textcolor{black}{OT detection by the \miniG{} telescopes} {is} strongly unfavored. For completeness we draw the same estimates for all the GW \textcolor{black}{alerts we followed-up} and summarize the results in Table \ref{tab:res_ser_assoc}.
\begin{table}[h!]
	\caption{{Comparison study between the number of OTs we may expect to detect during our follow-up campaign and those we actually detected. For each GW event, no OT has been found in agreement with our expectations. As a consequence, one OT detection would immediately lead to a strong probability of association with the real GW merger events (G268556 and G288732).}}
	\label{tab:res_ser_assoc}
	\centering
	\begin{tabular}{ll}
		\hline\hline\noalign{\smallskip}
		GW & $\mathrm{N^{ser.}_{OT,GW}}$ /   \\
		event &(OT detected) \\
		\hline
		\\
		G268556  & $\mathrm{2.6\times 10^{-2}}$ / (0) \\ 
		G270580  & $\mathrm{1.6\times 10^{-1}}$ / (0) \\
		G274296  & $\mathrm{3.5\times 10^{-2}}$ / (0) \\
		G275404  & $\mathrm{6.8\times 10^{-3}}$ / (0) \\
		G275697  & $\mathrm{7.7\times 10^{-3}}$ / (0) \\
		G277583  & $\mathrm{1.5\times 10^{-1}}$ / (0) \\
		G284239  & $\mathrm{4.4\times 10^{-2}}$ / (0) \\
		G288732  & $\mathrm{2.6\times 10^{-2}}$ / (0) \\
		\noalign{\smallskip}\hline
	\end{tabular}
\end{table}
We tentatively set an upper limit (U.L.) on the optical flux of GW170104 during our period of observations but under the hypothesis that the event was located in the portion of the sky we have monitored. This 3$\sigma$ U.L., \textcolor{black}{lying in the range} $m_R \in [12.3 - 11.4]$, \textcolor{black}{varies from a grid to an other one as} the sky brightness can significantly \textcolor{black}{change}. For GW170608, the limiting magnitude of our images are less stringent because of a cloudy sky. \textcolor{black}{The optical flux upper limit of GW176008 finally lie in the range} $m_R \in [10.9 - 9.9]$, again assuming that the event was localized in our images.\\

\section{Towards the next LIGO-Virgo O3 run}
\label{sec:O3}
The next GW scientific run on April 2019 (O3) promises to be prolific in terms of {the number} of GW detections that will need extensive electromagnetic follow-up campaigns too. Thanks to the sensitivity improvement of the LIGO-Virgo detectors, one can expect, in the most optimistic scenario, one BNS merger per month and most likely few BBH mergers per week. The {localization uncertainties} of the GW O3 {events} will be largely reduced due to the combination of the LIGO-Virgo detectors with a median localization \textcolor{black}{region comprised in the range} 120-170 deg$^2$ within the 90$\%$ confidence level contours for LIGO only\footnote{See the LIGO/Virgo prospects for the O3 run here \url{https://emfollow.docs.ligo.org/userguide/capabilities.html$\#$livingreview} and the associated references.}. 
Despite such {significant improvement of the localizations}, the need for wide field {of view} telescopes will be still \textcolor{black}{crucial} for some events. Furthermore, according to the expected high GW alert rate, the availability of world-wide networks of telescopes dedicated to the electromagnetic follow-up of the GW events will be \textcolor{black}{a key factor to make the O3 run a scientific success as O2 was}. 

\subsection{From the \miniG{} to the GWAC system}

\begin{figure*}[h!]
	\centering
	\includegraphics[trim = 0 0 160 30,clip=true, scale = 0.14]{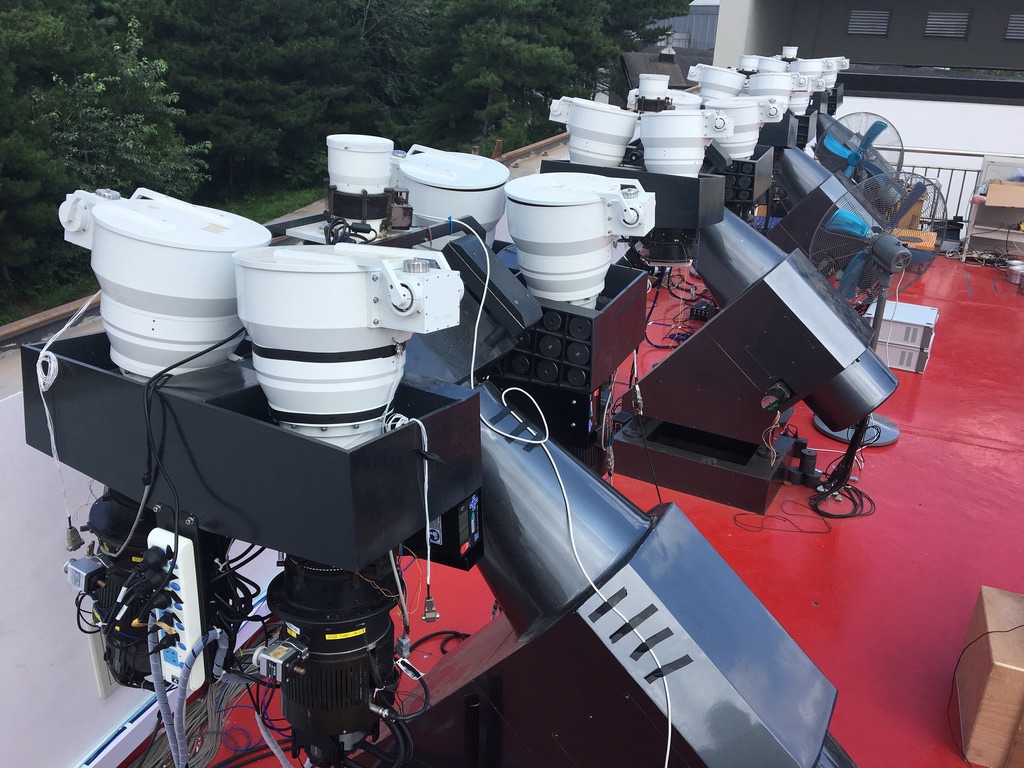}
	\includegraphics[trim= 0 0 0 0, clip=true,scale = 0.135]{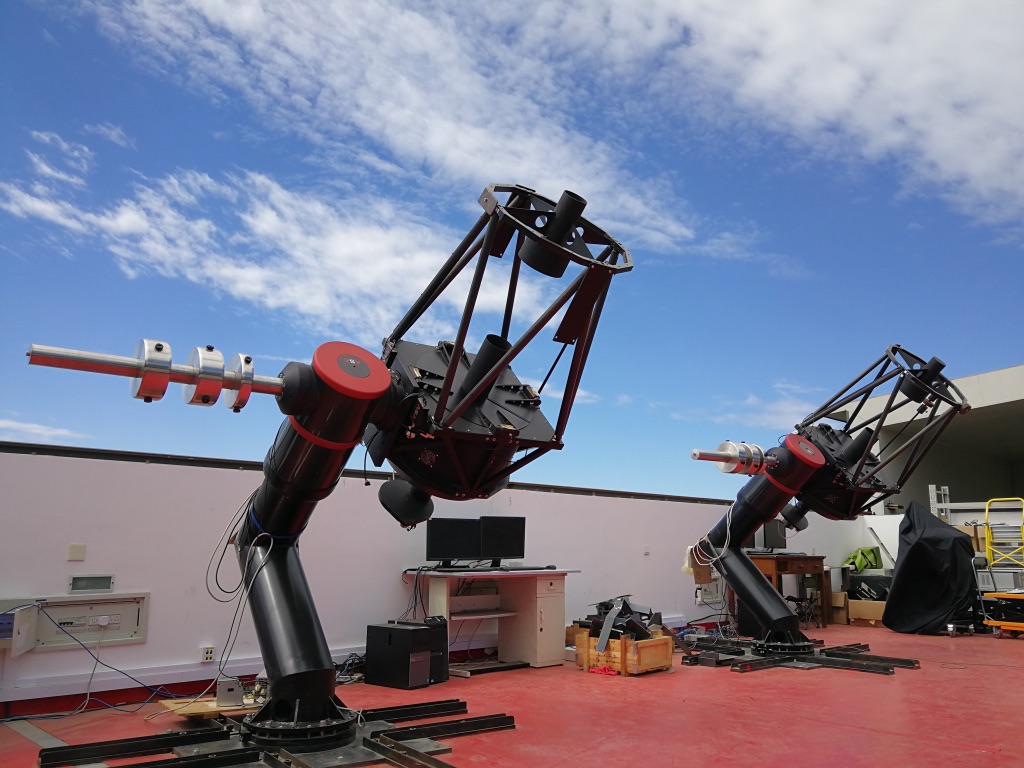}
	\includegraphics[trim= 0 0 0 0, clip=true,scale = 0.135]{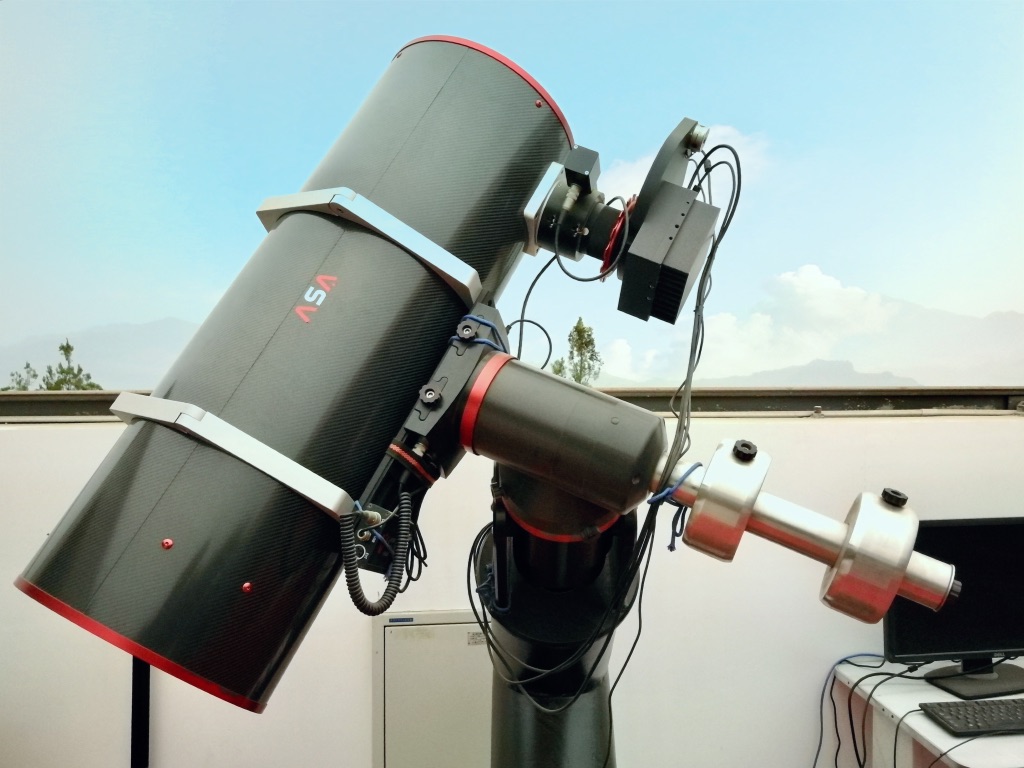}
	
	\caption{({\it Left}) The GWAC observation system in its test bench at the Xinglong observatory. For O3, it will be composed of 4 mounts. The total FoV of such configuration is about 2000 sq.deg. The imaging time resolution is 15 sec for single frames. ({\it Center}) Two GWAC-F60 telescopes (60cm) are used to quickly confirm GWAC optical transients and perform deeper follow-up observations if needed. ({\it Right}) A 30 cm telescope (GWAC-F30) will be also used during the O3 campaign (FoV = {1.8$^o \times$ 1.8$^o$}).}
	\label{Fig:gwac}
\end{figure*}

Since the end of 2017, \miniG{} have been totally replaced by the nominal design of the \GWAC{} telescopes and are no longer used.
Each \GWAC{} mount is equipped with 5 cameras (4 x JFoV camera: 4k x 4k CCD E2V camera with an aperture of 180 mm each + 1 FFoV: 3k x 3k CCD camera with an aperture of 35 mm), see Figure \ref{Fig:gwac}. With such a system, each mount will have a field of view of  about $25^\circ \times 25^\circ$ ($\sim$500 square degrees) with an optical flux coverage extended from V $\sim$ 6 magnitude up to 16 magnitude\footnote{This sensitivity is reached in a dark night for 10 seconds of exposure.} in the visible domain $\mathrm{\lambda \in}$ [500-850 nm]. As for \miniG{}, an image cadence of 15 seconds is set. For the O3 run, four \GWAC{} mounts will be available at the Xinglong Observatory\footnote{At completion, the \GWAC{} network system will be composed of a set of 10 mounts located in China and 10 others located out of China (the second site is still under discussion).}. {We summarize, in Table \ref{tab:GWAC_comp}, some parameters of the GWAC telescopes and compare them with those of the \miniG{} telescopes to highlight the improvements. The major improvements are the increase of the \GWAC{} sensitivity and {the} angular resolution {compared with} the \miniG{} system.}
\begin{table}[h!]
	\caption{{Comparison between some parameters of \miniG{} and \GWAC{}.}}
	\label{tab:GWAC_comp}
	\scriptsize
	\centering
	\begin{tabular}{llll}
		\hline\hline\noalign{\smallskip}
		Parameter & \miniG{} & \GWAC{} & \GWAC{}  \\
		          &  value   &  value  & improvement factor\\
		\hline
		\\
		Network FoV (sq.deg) & 5000 & 5000 & 1 \\
		Tel. diameter (cm) & 7.0 & 18 (JFOV) & $\sim$ 2.5\\
		Pixel size ($\mu$m) & 12 & 13 & $\sim$ 1 \\
		Pixel scale (arcsec) & 29.5 & 11.7 & 2.5\\
		Readout noise (e$^-$) & 10 & 14 & 0.7 \\
		FWHM (center) & 1.2 & 1.5 & 1.25 \\
		$\mathrm{R_{lim}}$ (mag/single frame) & 12 & 16 & $\sim$ 40 \\ 
		                                      &    &    & (in flux sensitivity)\\
		\noalign{\smallskip}\hline
	\end{tabular}

\end{table}
In association with the \GWAC{} telescopes, our two fully robotized 60 cm telescopes (GWAC-F60A/B) will be also used to automatically confirm the \textcolor{black}{genuineness of the} \GWAC{} OT candidates with a localization accuracy of the source of $\sigma\sim 1$ arcsec. They will also provide multi-wavelength (Johnson UBVRI) observations of the galaxies targeted in the GW probability skymaps. Finally, the \GWAC{} system will be completed by the GWAC-F30 robotic telescope (30 cm) \textcolor{black}{operated} with a substantial field of view of {1.8$^o \times$ 1.8$^o$} using different filters (Johnson UBVRI). 
As a whole, this \GWAC{} system offer multiple capabilities of observations and strategies for the optical follow-up of the gravitational wave alerts.

\subsubsection{Real-time stacking analysis and search for slow transient}
\begin{figure*}[h!]
	
	\begin{minipage}{1.0\textwidth}
		\centering
		\includegraphics[trim= 0 80 200 100, clip=true,scale=0.5]{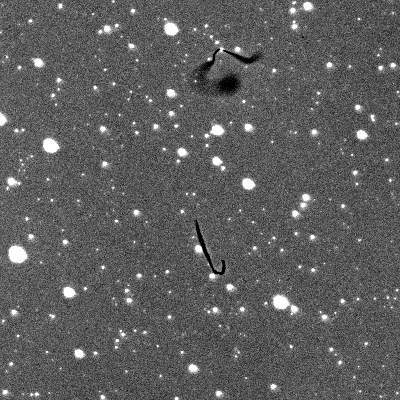}
		\includegraphics[trim= 0 80 200 100, clip=true,scale=0.5]{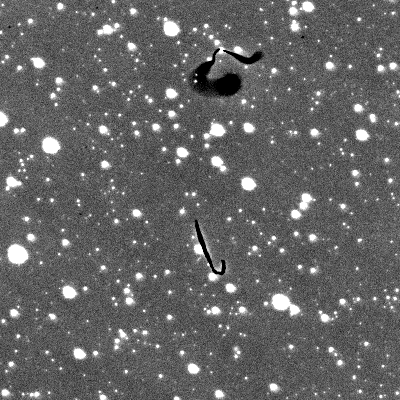}
		\includegraphics[trim= 0 80 200 100, clip=true,scale=0.5]{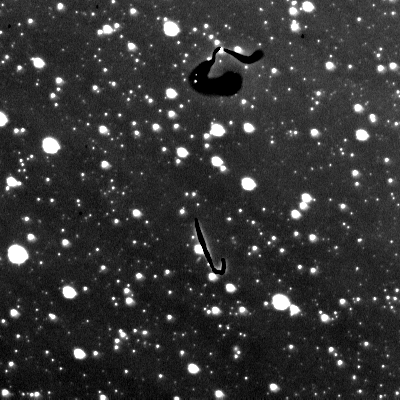}
		\includegraphics[trim= 0 80 200 100, clip=true,scale=0.5]{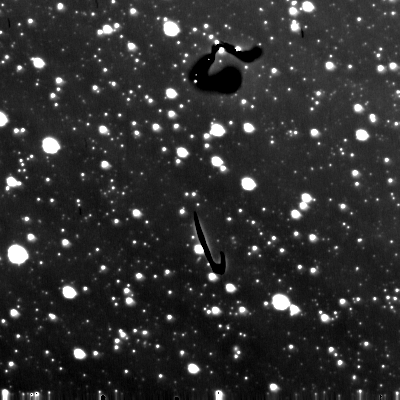}
	\end{minipage}
	\begin{minipage}{1.0\textwidth}
		\centering
		\includegraphics[width = 0.4\columnwidth]{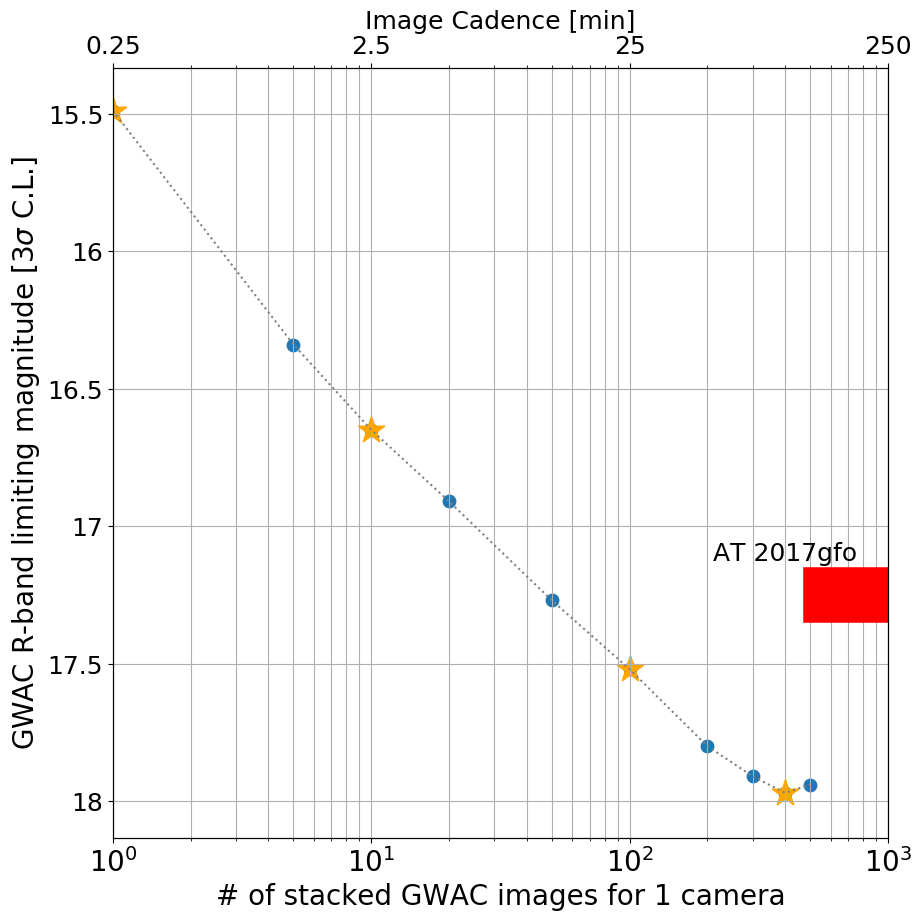}
	\end{minipage}
	\caption{(Top) Series of stacked GWAC sub-images using N = 1 (single image), 10, 100 and 400 images, respectively from left to right. From the left to the right images the limiting magnitude goes from R = 15.49 to R = 17.97 (calibrated with the USNO B1.0 R2mag stars). (Bottom ) Limiting magnitudes of GWAC (3$\sigma$ confidence level) as function of the number of stacked images. The orange stars represent the limiting magnitude of the GWAC images shown above. The kilonova (AT 2017gfo) associated with the GW170817 event is shown assuming a minimum typical timescale of 10 hours for the optical emission. Our stacking analysis would allow us to reach the detection threshold for such kind of event.}
	\label{fig:GWAC_stacking}
\end{figure*}

Once data \textcolor{black}{will be} taken, we will conduct a stacking analysis of our images to reach a maximum sensitivity of V$\sim$18 (a gain of six magnitude with respect to the \miniG{} system) in a time-resolution of several hours while keeping a high imaging quality as shown in Figure \ref{fig:GWAC_stacking}. 
This kind of set-up is built to search for moderately slow fading and faint transients having low signal-to-noise ratios in our single images. The stacking analysis of \GWAC{} images would permit to reach the detection threshold of the kilonova optical emisison near its maximum brightness if such events are as close and bright as AT 2017gfo, {the kilonova optical counterpart of the BNS merger GW 170817}, (m$_{R,peak}\sim 17.2$).
The discovery potential of GRB optical afterglow emission is also highly enhanced with such increase of our sensitivity. However, in the case of the GRB afterglows, the geometry of the emission can significantly affect our detection capability, whether the electromagnetic emission is isotropically radiated or through a narrow jet. If a jet is involved, its viewing angle will also play a significant role. If it is \textcolor{black}{seen} largely off-axis compared to our line of sight, the electromagnetic flux we may receive will be strongly reduced and delayed, \textcolor{black}{hence disfavoring an optical detection by our telescopes. On the contrary, for a jetted emission seen on-axis at the BNS distance range of LIGO-Virgo for O3 (120Mpc - 60Mpc), we will very likely detect the optical emission that is expected to be significantly brighter than the GWAC sensitivity (R = 16 mag) at early time post merger.}

\subsubsection{Automatic and quick classification of the transient candidates}
A key challenge of the wide field {of view} telescopes is to be able to quickly {identify and classify} the numerous transient sources they detect each night. Despite the field of view of {the} \miniG{} {telescopes} was very large, their limiting {sensitivities} prevented them from detecting a huge number of optical transients every night (few dozens of OT {candidates} per mount). Therefore, it was still possible to fully involved humans in the loop of the source classification. For \GWAC{}, it will be no longer the case as the sensitivity of each mount is significantly increased and especially considering the real-time stacking analysis. Typically, in one dark night, the \GWAC{} detection pipeline can be triggered (at the very basic level of OT1) hundreds of times using only single images and the cameras of one mount. {As explained in \ref{sec:data_proc}, the} preliminary sample of OT candidates is usually composed of artifacts and {possibly} few genuine astrophysical sources. A new method of OT classification has been developed in the frame work of the \GWAC{} data processing pipeline based on a machine learning approach. This new classification method, that will be described in detail in a separate paper, will use Convolutional Neural Networks (CNN). This approach is now widely used for telescopes {having} wide field of {views} {\citep[e.g.][]{Gieseke17,Sanchez18,Mahabal19,Jia19} and is particularly efficient in detecting bogus in images such as cosmic-rays, hot pixels, etc. (the category D of our own classification ranking, see \ref{sec:OT_class}) which constitute the major fraction of our false detections at the OT1 level. The goal is to filter out around 95\% of the false positives detected in our \textit{OT1 candidate} sample.} It is crucial for such telescopes in order to be efficient in detecting "the good ones" and to ensure that our optical transient candidates will be of a great interest for the astronomical community when we will release public \GWAC{} alerts.

\subsubsection{The first training of the SVOM ground follow-up system.}
In 2021, the SVOM mission will be endowed with a network of ground optical/NIR telescopes devoted to the follow-up of the SVOM triggers or ToO triggers approved by the SVOM Collaboration \citep{SVOM16}. At completion, this ground segment should be composed of the SVOM/COLIBRI telescope located at the Observatory of San Pedro M\'artir (Mexico), a set of ten GWAC mounts located out of China (the location is still under discussion) and some telescopes located in China: ten GWAC mounts, two GWAC-F60, one GWAC-F30 and the C-GFT telescope (1.2m). For the O3 run, only the chinese part of the SVOM segment will be available with four operational GWAC telescopes and also including the C-GFT telescope. The goal of this chinese network is to pave the GW skymap in the most efficient way by combining different observational strategies such as tiling observations of the GW skymap or galaxy targeting. This strategies will take into account the individual characteristics of our telescopes that will be connected to the SVOM chinese science center (CSC) for O3 at the National Astronomical Observatories of China (NAOC), CAS. The CSC will be in charge of collecting all the observational results and producing the public reports. This centralized database system will allow us to adapt our strategy almost in real time depending whether we need to explore new fields, make some revisit observations or confirm optical transient candidates with multi-band photometric observations. With such system, we will provide, as fast as possible and publicly through the GCN network, the list of the most interesting OT candidates we have found: the so-called \textit{OT4 candidates} according to our internal labeling system described above. In order to better characterize these promising OTs, based on their temporal behavior and their color evolution, we will conduct spectroscopic follow-up observations using the 2.16m telescope at the Xinglong Observatory and the 2.4m telescope at the Lijiang station of the Yunnan Observatory. Note that we can also perform deep color photometry with such telescopes with a limiting magnitude B/V/R $\sim$ 22 for 10 minute exposure time (under an airmass = 1.3) with the BFOSC instrument mounted on the 2.16m telescope \citep{Fan16}. For the same exposure time, we can reach a R $\sim 24$ limiting magnitude with the 2.4 m telescopes. During the O2 run, we performed such deep follow-up observations with the 2.16m telescope at Xinglong for an optical transient detected by Swift/UVOT related to the GW trigger G299232 \citep{Meng17}. We could not detect this transient down to a magnitude r $\sim$ 22 confirming its fading behavior compared to the Swift/UVOT data and consistent with observations performed by other teams. This example shows how these {moderately} large telescopes will allow us to extend our follow-up capabilities for faint sources (r $<22$) to possibly detect sources similar to the GW170817/AT 2017gfo kilonova \citep{Villar17} days after the merger event.

\section{Conclusion and Perspectives}
\label{sec:Conclusion}

The O2 GW observational campaign has opened a new window to study the extreme objects in the Universe. It helped us to validate the capability of the \miniG{} telescope network as being a fast follow-up system dedicated to the multi-messenger astronomy. So far, our O2 observation campaign represents the largest coverage of the GW sky localization areas made by optical telescopes in short latencies. No credible optical transient was found in our images which we attribute to two main reasons. First, the confirmed GW events we have followed-up, were all originating from BBH mergers from which an electromagnetic emission is highly uncertain. Secondly, the sensitivity of the \miniG{} telescopes ($m_R = 12$) was too low to detect faint transient sources such as the kilonova emission like the one observed for \textcolor{black}{GW170817/AT2017gfo} or any GRB afterglow emission. Based on this experience, we have presented our new plan for the upcoming O3 run. We showed that the improvement of our \textcolor{black}{observational} capabilities by combining both a migration from the \miniG{} to the \GWAC{} system, with a much higher sensitivity in the visible domain, and the extension of our network will permit us to be more competitive in our searches for optical counterparts from GW events, especially those emerging from the BNS mergers. The O3 run will be also a unique opportunity to build the first blocks of the ground follow-up system of the future SVOM mission {that embedded the GWAC system}.

\section*{Acknowledgements}
	This work is supported by the National Natural Science Foundation of China (Grant No. 11533003, 11673006,  U1331202),  Guangxi Science Foundation (2016GXNSFFA380006, AD17129006, 2018GXNSFGA281007), as well as the Strategic Priority Research Program of the Chinese Academy of SciencesGrant No.XDB23040000 and the Strategic Pionner Program on Space Science, Chinese Academy of Sciences, Grant No.XDA15052600, and D.Turpin acknowledges the financial support from the Chinese Academy of Sciences (CAS) PIFI post-doctoral fellowship program (program C). S. Antier, B. Cordier and C. Lachaud acknowledge the financial support of the UnivEarthS Labex program at Sorbonne Paris Cit\'e (ANR-10-LABX-0023 and ANR-11-IDEX-0005-02).

\appendix

\section*{Appendices}
\section{The mini-GWAC follow-up observations of eight O2 GW alerts.}
\begin{figure*}[h!]
	\begin{minipage}{1.0\textwidth}
		\centering
		\includegraphics[width=0.45\columnwidth]{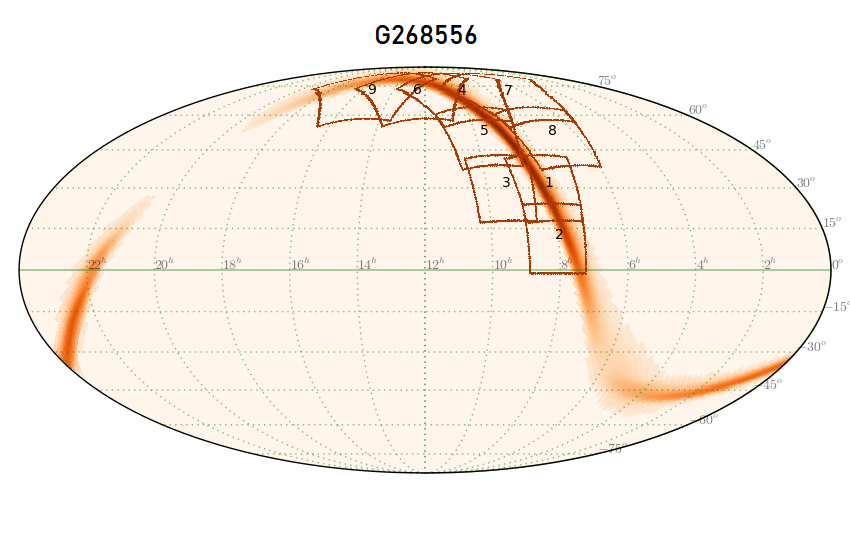}
		\includegraphics[width=0.45\columnwidth]{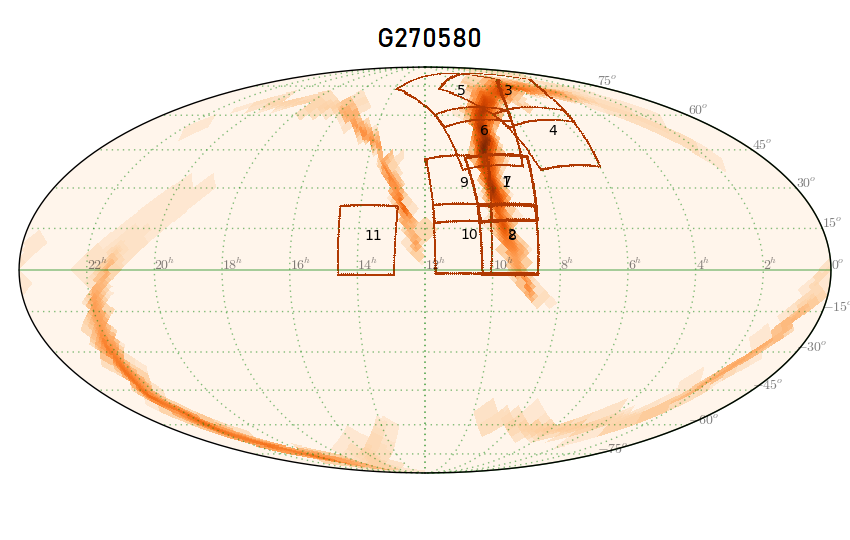}
	\end{minipage}
	\begin{minipage}{1.0\textwidth}
		\centering
		\includegraphics[width=0.45\textwidth]{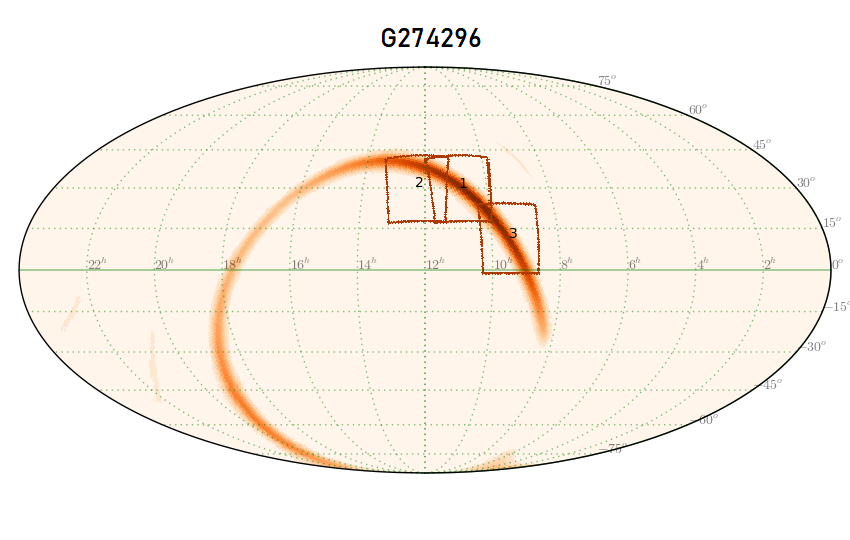}
		\includegraphics[width=0.45\textwidth]{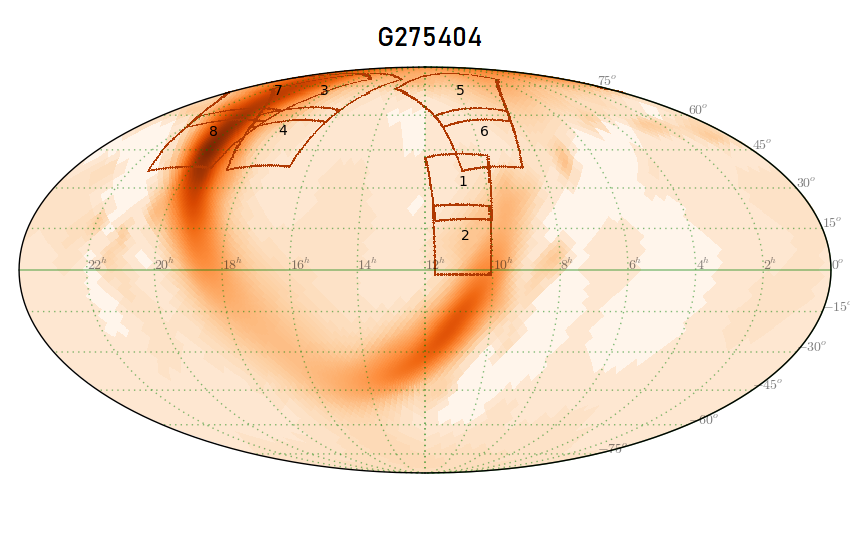}
	\end{minipage}
	\begin{minipage}{1.0\textwidth}
		\centering
		\includegraphics[width=0.45\textwidth]{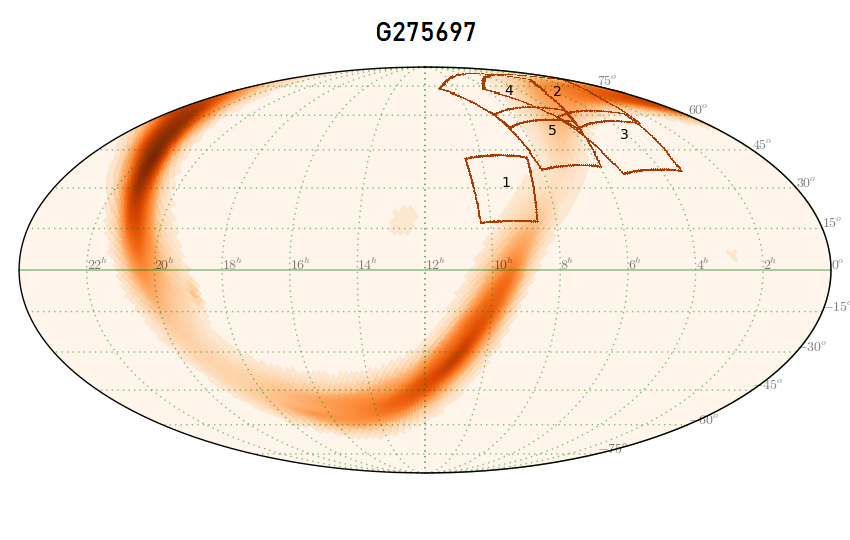}
		\includegraphics[width=0.45\textwidth]{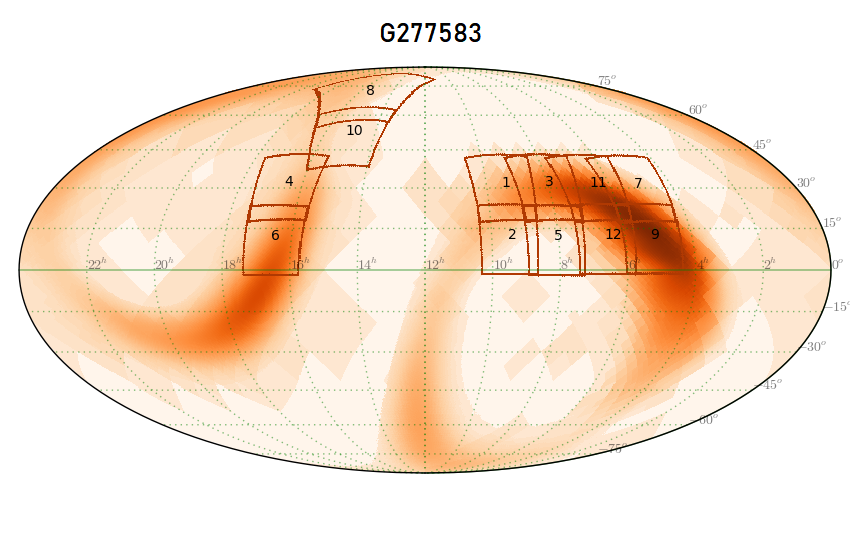}
	\end{minipage}
	\begin{minipage}{1.0\textwidth}
		\centering
		\includegraphics[width=0.45\textwidth]{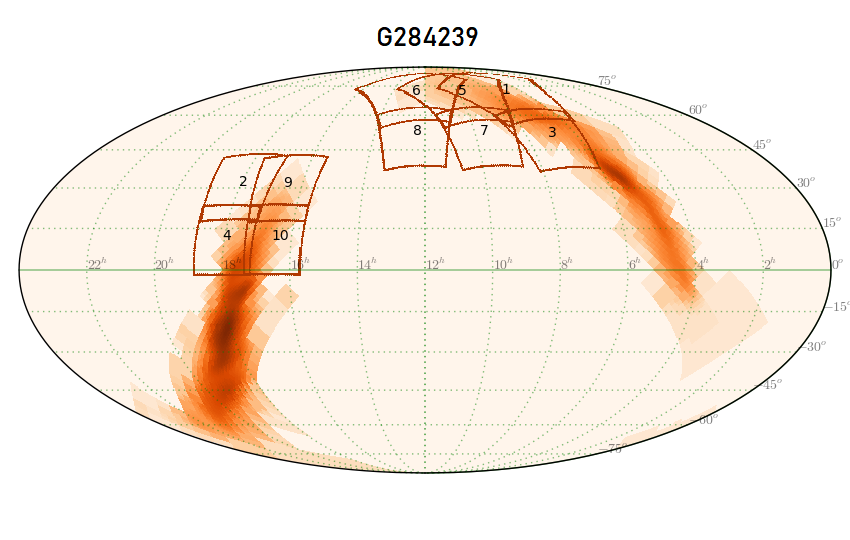}
		\includegraphics[width=0.45\textwidth]{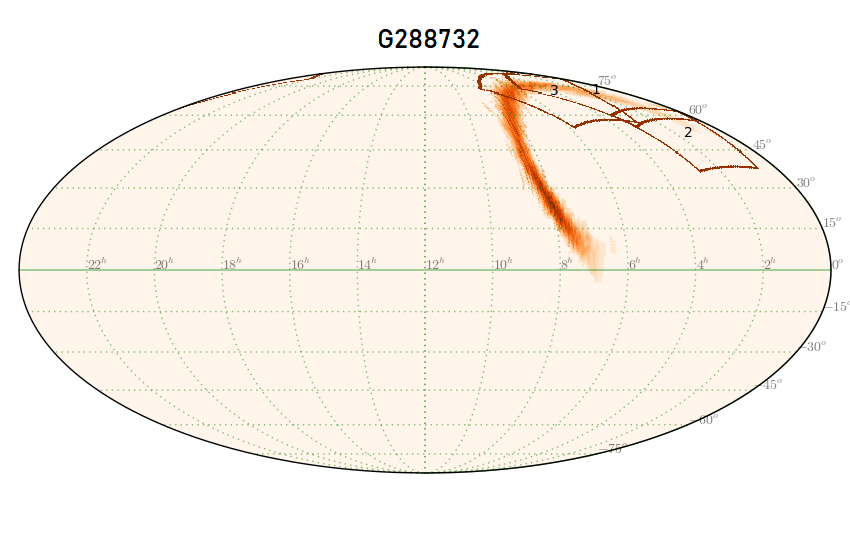}
	\end{minipage}
	\caption{The bayesian probability skymaps of the eight gravitational wave events we followed-up during the O2 run. Our observation grids are shown with the red squares, each of them are identified with a grid ID. All these grids were not necessarily scheduled at the same period because of observational constraints but it shows how we paved the GW error boxes all along our periods of observation.}
	\label{Fig:GW_skymaps}
\end{figure*}
\clearpage
\newpage
\section{The Log. tables of the \miniG{} observation performed for eight GW events during the O2 LIGO/Virgo run.}
\label{Appendix}

\begin{table*}[h!]
	\caption{The observation logs of the mini-GWAC follow-up of G268556 (trig. date: 2017-01-04 10:11:58).}
	\label{tab:G268556_obs}
	\scriptsize
	\begin{center}
		\begin{tabular}{ccclcccll}
			\hline
			mini-GWAC & $\mathrm{T_{start}}$ & $\mathrm{T_{end}}$  & ~~~~mid time & center RA & center dec & $\mathrm{P_{GW,cov}}$ & $\mathrm{N_{im}}$ / $\mathrm{N_{OT2}}$  & $\mathrm{M_{R,OT2}}$ \\
			grid / cam ID & 2017-01-04 & 2017-01-04  & ~~~~~ (hour)  & (h:m:s) & (deg:m:s) &  & & [min - max] \\
			\hline
			1 / C1 & 12:30:41.1 & 13:49:41.5 & $\mathrm{T_{GW}+2.9704}$ &07:46:49.578 & +29:35:33.46 & 18.5$\%$ & 316 / 50 & [12.3 - 8.7] \\
			2 / C2 & 12:30:41.1 & 13:49:49.5 & $\mathrm{T_{GW}+2.9715}$ &07:48:54.239 & +10:34:56.09 & 13.4$\%$ & 317 / 36 & [11.9 - 8.2] \\
			3 / C1 & 13:50:29.3 & 15:14:52.1 & $\mathrm{T_{GW}+4.3452}$ & 09:10:51.599 & +29:36:54.60 & 3.1$\%$ & 338 / 0 & -- \\
			7 / C5 & 14:55:58.2 & 17:57:22.7 & $\mathrm{T_{GW}+6.2451}$ & 06:34:42.357 & +69:28:01.79 & 3.9$\%$ & 726 / 0 & -- \\
			8 / C6 & 14:56:10.4 & 17:57:35.7 & $\mathrm{T_{GW}+6.2486}$ & 06:40:16.529 & +50:28:28.89 & 0.3$\%$ & 726 / 0 & -- \\
			6 / C3 & 16:21:28.7 & 17:57:31.9 & $\mathrm{T_{GW}+6.9590}$ & 11:52:01.006 & +70:06:03.83 & 11.0$\%$ & 384 / 0 & -- \\
			4 / C1 & 19:14:27.3 & 22:39:37.7 & $\mathrm{T_{GW}+10.7513}$ & 09:17:21.644 & +69:37:03.40 & 17.7$\%$ & 821 / 142 & [11.4 - 6.8] \\
			5 / C2 & 19:14:27.3 & 22:39:25.3 & $\mathrm{T_{GW}+10.7495}$ & 09:21:25.794 & +50:35:59.26 & 16.7$\%$ & 820 / 1 & 9.9 \\
			6 / C5 & 19:14:32.9 & 22:39:31.9 & $\mathrm{T_{GW}+10.7512}$ & 11:52:01.006 & +70:06:03.83 & 11.0$\%$ & 820 / 159 & [11.4 - 6.8] \\
			1 / C3 & 19:14:39.9 & 21:17:22.8 & $\mathrm{T_{GW}+10.0676}$ & 07:46:49.578 & +29:35:33.46 & 18.5$\%$ & 490 / 2 & [11.1 - 10.5] \\
			2 / C4 & 19:14:39.9 & 21:17:36.8 & $\mathrm{T_{GW}+10.0695}$ & 07:48:54.239 & +10:34:56.09 & 13.4$\%$ & 492 / 24 & [11.1 - 8.3] \\     
			9 / C7 & 19:14:55.3 & 22:39:18.3 & $\mathrm{T_{GW}+10.7524}$ & 14:34:10.239 & +70:01:52.84 & 7.1$\%$ & 818 / 110 & [11.4 - 6.8] \\
			\hline
		\end{tabular}
	\end{center}
	\textit{Note:} The time of each observation is given in UTC. $\mathrm{T_{start}}$ and $\mathrm{T_{end}}$ correspond to the interval time during which the mini-GWAC telescopes were taking images (with a cadence of 15s). The mid time of the whole mini-GWAC observations is computed in the interval [$\mathrm{T_{start}}-\mathrm{T_{end}}$]. The RA and dec coordinates of the images stand for the center of each image (FoV $\sim 20^\circ \times 40^\circ$). The number of images as well as the number of optical transient candidates detected during the whole observation period are given for information with $\mathrm{N_{im}}$ and $\mathrm{N_{OT2}}$, respectively. Note that several OT candidates might be detected by different cameras as there are significant overlaps between the observed fields. Finally, $\mathrm{M_{R,OT2}}$ corresponds to the range of magnitudes where the OT candidates were found in single images (unfiltered calibrated with R/Johnson).
\end{table*}

\begin{table*}[h!]
	\centering
	\caption{The observation logs of the mini-GWAC follow-up of G270580 ( trig. date: 2017-01-20 12:30:59.35). Same caption as for table \ref{tab:G268556_obs}.}
	\label{tab:G270580_obs}
	\scriptsize
	\begin{tabular}{ccclcccll}
		\hline
		mini-GWAC & $\mathrm{T_{start}}$ & $\mathrm{T_{end}}$  & ~~~~mid time & center RA & center dec & $\mathrm{P_{GW,cov}}$ & $\mathrm{N_{im}}$ / $\mathrm{N_{OT2}}$  & $\mathrm{M_{R,OT2}}$  \\
		grid / cam ID & 2017-01-20 & 2017-01-20  & ~~~~~ (hour)  & (h:m:s) & (deg:m:s) &  & & [min - max]  \\
		\hline
		1 / C1 & 12:50:28.3 & 14:15:07.6 & $\mathrm{T_{GW}+1.0302}$ & 09:10:23.301 & +29:35:57.71 & 16.2$\%$ & 339 / 1 & 8.6 \\
		2 / C2 & 12:50:28.3 & 22:14:58.6 & $\mathrm{T_{GW}+5.0289}$ & 09:12:26.259 & +10:35:26.10 & 8.3$\%$ & 2258 / 0 & -- \\
		3 / C5 & 13:50:51.4 & 19:47:59.1 & $\mathrm{T_{GW}+4.3072}$ & 06:36:32.060 & +69:30:22.27 & 12.7$\%$ & 1429 / 20 & [11.7 - 9.6] \\
		4 / C6 & 13:50:51.4 & 19:48:01.2 & $\mathrm{T_{GW}+4.3072}$ & 06:42:03.137 & +50:31:15.23 & 0.2$\%$ & 1429 / 0 & -- \\
		5 / C1 & 14:15:35.4 & 22:19:45.5 & $\mathrm{T_{GW}+5.7781}$ & 09:17:48.639 & +69:36:41.75 & 12.9$\%$ & 1937 / 6 & [11.8 - 10.2] \\
		6 / C2 & 14:15:35.4 & 22:19:48.5 & $\mathrm{T_{GW}+5.7785}$ & 09:21:16.884 & +50:34:54.73 & 23.0$\%$ & 1937 / 2 & [10.3 - 10.2]\\
		1 / C3 & 14:16:01.4 & 21:24:39.2 & $\mathrm{T_{GW}+5.3224}$ & 09:08:57.610 & +30:01:54.06 & 16.4$\%$ & 1715 / 4 & [11.8 - 8.4] \\
		2 / C4 & 14:16:01.4 & 21:25:03.9 & $\mathrm{T_{GW}+5.3259}$ & 09:13:11.448 & +09:57:30.75 & 8.0$\%$ & 1716 / 3 & [11.5 - 11.1] \\
		9 / C5 & 19:49:04.7 & 21:36:13.2 & $\mathrm{T_{GW}+8.1943}$ & 10:34:34.322 & +29:30:16.35 & 5.3$\%$ & 429 / 1 & 8.4\\
		10 / C6 & 19:49:27.9 & 22:19:41.5 & $\mathrm{T_{GW}+8.5598}$ & 10:37:18.165 & +10:30:56.78 & 0.1$\%$ & 601 / 0 & -- \\
		11 / C4 & 21:32:19.1 & 22:19:53.1 & $\mathrm{T_{GW}+9.4185}$ & 13:29:12.042 & +10:00:17.26 & 0.1$\%$ & 190 / 0 & -- \\
		\hline
	\end{tabular}
\end{table*}

\begin{table*}[h!]
	\centering
	\caption{The observation logs of the mini-GWAC follow-up of G274296 ( trig. date: 2017-02-17 06:05:55.05). Same caption as for table \ref{tab:G268556_obs}. $^\dagger$ For this set of observations the corresponding date is 2017-02-18.}
	\label{tab:G274296_obs}
	\scriptsize
	\begin{tabular}{ccclcccll}
		\hline
		mini-GWAC & $\mathrm{T_{start}}$ & $\mathrm{T_{end}}$  & ~~~~mid time & center RA & center dec & $\mathrm{P_{GW,cov}}$ & $\mathrm{N_{im}}$ / $\mathrm{N_{OT2}}$  & $\mathrm{M_{R,OT2}}$  \\
		grid / cam ID & 2017-02-17 & 2017-02-17  & ~~~~~ (hour)  & (h:m:s) & (deg:m:s) &  & & [min - max] \\
		\hline
		1 / C1 & 12:20:29.0 & 13:45:04.7 & $\mathrm{T_{GW}+6.1144}$ & 10:34:48.326 & +29:29:08.60 & 32.0$\%$ & 338 / 4 & [12.2 - 8.5] \\
		2 / C1 & 13:45:30.2 & 17:12:33.6 & $\mathrm{T_{GW}+8.5519}$ & 11:58:53.431 & +29:29:28.69 & 17.4$\%$ & 828 / 1 & 9.6 \\
		3 / C6$^\dagger$ & 10:53:52.3 & 12:57:00.8 & $\mathrm{T_{GW}+28.9920}$ & 09:12:10.933 & +10:39:50.19 & 27.7$\%$ & 493 / 0& -- \\
		
		\hline
	\end{tabular}
\end{table*}

\begin{table*}[h!]
	\centering
	\caption{The observation logs of the mini-GWAC follow-up of G275404 ( trig. date: 2017-02-25 18:30:21). Same caption as for table \ref{tab:G268556_obs}.}
	\label{tab:G275404_obs}
	\scriptsize
	\begin{tabular}{ccclcccll}
		\hline
		mini-GWAC & $\mathrm{T_{start}}$ & $\mathrm{T_{end}}$  & ~~~~mid time & center RA & center dec & $\mathrm{P_{GW,cov}}$ & $\mathrm{N_{im}}$ / $\mathrm{N_{OT2}}$  & $\mathrm{M_{R,OT2}}$  \\
		grid / cam ID & 2017-02-25 & 2017-02-25  & ~~~~~ (hour)  & (h:m:s) & (deg:m:s) &  & & [min - max] \\
		\hline
		5 / C7 & 13:01:04.2 & 21:37:38.1 & $\mathrm{T_{GW}-1.1832}$ & 09:21:25.5 & +69:40:01 & 1.4$\%$ & 2066 / 0 & -- \\
		6 / C8 & 13:01:04.2 & 21:37:39.7 & $\mathrm{T_{GW}-1.1831}$ & 09:23:01.6 & +50:00:25 & 0.4$\%$ & 2066 / 1 & 12.0 \\
		1 / C3 & 19:23:51.9 & 20:41:04.7 & $\mathrm{T_{GW}+1.5354}$ & 10:33:59.6 & +30:12:22 & 0.5$\%$ & 309 / 2 & [11.6 - 9.2] \\
		2 / C4 & 19:23:51.9 & 20:41:06.2 & $\mathrm{T_{GW}+1.5356}$ & 10:38:05.6 & +10:07:57 & 1.8$\%$ & 309 / 50 & [12.1 - 5.3]\\
		3 / C5 & 19:23:42.2 & 22:13:19.6 & $\mathrm{T_{GW}+2.3027}$ & 17:18:04.0 & +69:28:00 & 6.9$\%$ & 678 / 32 & [12.2 - 10.4] \\
		4 / C6 & 19:23:44.7 & 22:13:38.0 & $\mathrm{T_{GW}+2.3057}$ & 17:21:47.1 & +50:28:32 & 2.0$\%$ & 680 / 4 & [11.9 - 11.8] \\
		7 / C7 & 21:39:38.3 & 22:13:28.5 & $\mathrm{T_{GW}+3.4368}$ & 20:00:36.6 & +69:38:42 & 12.5$\%$ & 135 / 0 & -- \\
		8 / C8 & 21:39:38.3 & 22:13:25.5 & $\mathrm{T_{GW}+3.4364}$ & 20:01:12.7 & +50:00:21 & 16.4$\%$ & 135 / 0 & -- \\
		\hline
	\end{tabular}
\end{table*}

\begin{table*}[h!]
	\centering
	\caption{The observation logs of the mini-GWAC follow-up of G275697 ( trig. date: 2017-02-27 18:57:31). Same caption as for table \ref{tab:G268556_obs}.}
	\label{tab:G275697_obs}
	\scriptsize
	\begin{tabular}{ccclcccll}
		\hline
		mini-GWAC & $\mathrm{T_{start}}$ & $\mathrm{T_{end}}$  & ~~~~mid time & center RA & center dec & $\mathrm{P_{GW,cov}}$ & $\mathrm{N_{im}}$ / $\mathrm{N_{OT2}}$  & $\mathrm{M_{R,OT2}}$  \\
		grid / cam ID & 2017-03-01 & 2017-03-01  & ~~~~~ (hour)  & (h:m:s) & (deg:m:s) &  & & [min - max]  \\
		\hline
		1 / C1 & 10:55:43.4 & 18:11:04.6 & $\mathrm{T_{GW}+43.5981}$ & 09:10:04.5 & +29:30:47 & 0.3$\%$ & 1741 / 0 & -- \\
		2 / C3 & 10:55:26.9 & 14:04:55.7 & $\mathrm{T_{GW}+41.5445}$ & 03:52:34.0 & +68:53:23 & 5.0$\%$ & 758 / 0 & -- \\
		3 / C4 & 10:55:26.9 & 14:04:55.7 & $\mathrm{T_{GW}+41.5445}$ & 04:02:12.2 & +48:48:08 & 0.4$\%$ & 758 / 0 & -- \\
		4 / C5 & 10:55:24.5 & 17:44:07.8 & $\mathrm{T_{GW}+43.3709}$ & 06:34:55.5 & +69:32:01 & 1.5$\%$ & 1635 / 0 & -- \\
		5 / C6 & 10:55:24.5 & 17:44:07.8 & $\mathrm{T_{GW}+43.3709}$ & 06:40:15.5 & +50:32:35 & 1.3$\%$ & 1635 / 0 & -- \\
		\hline
	\end{tabular}
\end{table*}

\begin{table*}[h!]
	\centering
	\caption{The observation logs of the mini-GWAC follow-up of G277583 ( trig. date: 2017-03-13 22:40:09.59). Same caption as for table \ref{tab:G268556_obs}.}
	\label{tab:G277583_obs}
	\scriptsize
	\begin{tabular}{ccclcccll}
		\hline
		mini-GWAC & $\mathrm{T_{start}}$ & $\mathrm{T_{end}}$  & ~~~~mid time & center RA & center dec & $\mathrm{P_{GW,cov}}$ & $\mathrm{N_{im}}$ / $\mathrm{N_{OT2}}$  & $\mathrm{M_{R,OT2}}$  \\
		grid / cam ID & 2017-03-14 & 2017-03-14  & ~~~~~ (hour)  & (h:m:s) & (deg:m:s) &  & & [min - max] \\
		\hline
		9 / C6 & 11:10:11 & 13:33:39 & $\mathrm{T_{GW}+13.6959}$ & 04:58:00.5 & +10:28:12 & 19.2$\%$ & 574 / 35 & [11.5 - 7.6] \\
		1 / C1 & 11:10:29 & 17:59:02 & $\mathrm{T_{GW}+15.9100}$ & 09:10:29.8 & +29:50:17 & 2.4$\%$ & 1634 / 18 & [12.2 - 9.1] \\
		7 / C5 & 11:10:30 & 13:33:01 & $\mathrm{T_{GW}+13.6933}$ & 04:55:10.8 & +29:25:59 & 7.3$\%$ & 570 / 41 & [11.7 - 8.9]\\
		3 / C3 & 11:10:45 & 16:40:29 & $\mathrm{T_{GW}+15.2576}$ & 07:46:15.6 & +29:57:47 & 6.2$\%$ & 1319 / 16 & [12.0 - 6.8] \\
		5 / C4 & 11:10:45 & 16:40:01 & $\mathrm{T_{GW}+15.2537}$ & 07:50:25.4 & +09:54:53 & 1.1$\%$ & 1317 / 9 & [11.8 - 9.8] \\
		2 / C2 & 11:10:55 & 17:59:54 & $\mathrm{T_{GW}+15.9208}$ & 09:13:10.6 & +10:17:08 & 1.0$\%$ & 1635 / 4 & [11.4 - 9.9] \\
		12 / C8 & 12:45:01 & 15:56:16 & $\mathrm{T_{GW}+15.6747}$ & 06:20:02.3 & +10:21:24 & 8.2$\%$ & 765 / 52 & [11.8 - 9.3] \\
		11 / C7 & 12:45:41 & 14:56:37 & $\mathrm{T_{GW}+15.1832}$ & 06:18:37.7 & +29:46:06 & 11.5$\%$ & 524 / 22 & [11.6 - 8.9] \\
		10 / C6 & 13:34:09 & 21:30:00 & $\mathrm{T_{GW}+18.8653}$ & 14:40:07.7 & +50:29:58 & 0.4$\%$ & 1903 / 0 & -- \\
		8 / C5 & 13:34:31 & 21:30:00 & $\mathrm{T_{GW}+18.8683}$ & 14:36:21.1 & +69:30:06 & 0.8$\%$ & 1902 / 1 & 10.7\\
		6 / C4 & 16:50:38 & 21:30:00 & $\mathrm{T_{GW}+20.5026}$ & 16:18:18.2 & +09:59:04 & 5.3$\%$ & 1117 / 0 & -- \\
		4 / C3 & 16:50:46 & 21:30:00 & $\mathrm{T_{GW}+20.5037}$ & 16:14:18.2 & +30:03:36 & 1.5$\%$ & 1117 / 0 & -- \\
		\hline
	\end{tabular}
\end{table*}

\begin{table*}[h!]
	\centering
	\caption{The observation logs of the mini-GWAC follow-up of G284239 ( trig. date: 2017-05-02 22:26:07.91). Same caption as for table \ref{tab:G268556_obs}.}
	\label{tab:G284239_obs}
	\scriptsize
	\begin{tabular}{ccclcccll}
		\hline
		mini-GWAC & $\mathrm{T_{start}}$ & $\mathrm{T_{end}}$  & ~~mid time & center RA & center dec & $\mathrm{P_{GW,cov}}$ & $\mathrm{N_{im}}$ / $\mathrm{N_{OT2}}$  & $\mathrm{M_{R,OT2}}$ \\
		grid / cam ID & 2017-05-05 & 2017-05-05  & ~~~ (hour)  & (h:m:s) & (deg:m:s) &  &  & \\
		\hline
		5 / C5 & 12:10:29 & 17:09:26 & $\mathrm{T_{GW}+64.2304}$ & 09:15:43.9 & +69:29:34 & 3.2$\%$ & 1196 / 0 & --\\
		7 / C6 & 12:10:29 & 17:09:36 & $\mathrm{T_{GW}+64.2318}$ & 09:21:17.0 & +50:29:41 & 0.7$\%$ & 1196 / 0 & --\\
		9 / C7 & 12:11:12 & 20:09:24 & $\mathrm{T_{GW}+65.7361}$ & 16:15:56.4 & +29:39:59 & 1.7$\%$ & 1913 / 0 & --\\
		3 / C4 & 12:15:52 & 14:07:41 & $\mathrm{T_{GW}+62.7607}$ & 06:44:38.4 & +49:45:43 & 6.9$\%$ & 447 / 0 & --\\
		1 / C3 & 12:18:07 & 14:09:34 & $\mathrm{T_{GW}+62.7952}$ & 06:35:33.2 & +70:03:11 & 6.1$\%$ & 446 / 0 & --\\
		10 / C8 & 12:45:41 & 20:09:30 & $\mathrm{T_{GW}+66.0243}$ & 16:16:57.1 & +10:13:56 & 6.2$\%$ & 1775 / 30 & [11.8 - 10.3]\\
		4 / C4 & 14:15:38 & 20:09:05 & $\mathrm{T_{GW}+66.7704}$ & 17:45:33.4 & +10:00:59 & 5.2$\%$ & 447 / 17 & [11.8 - 9.9]\\
		2 / C3 & 14:17:02 & 20:09:33 & $\mathrm{T_{GW}+66.7860}$ & 17:41:40.8 & +30:05:48 & 0.3$\%$ & 1410 / 0 & --\\
		8 / C6 & 17:10:50 & 20:04:52 & $\mathrm{T_{GW}+68.1953}$ & 12:01:27.8 & +50:23:40 & $<0.1\%$ & 696 / 0 & --\\
		6 / C5 & 19:59:14 & 20:09:17 & $\mathrm{T_{GW}+69.6354}$ & 11:56:29.8 & +69:27:40 & 0.8$\%$ & 40 / 0 & --\\
		\hline
	\end{tabular}
\end{table*}

\begin{table*}[h!]
	\centering
	\caption{The observation logs of the mini-GWAC follow-up of G288732 ( trig. date: 2017-06-08 02:01:16.492). Same caption as for table \ref{tab:G268556_obs}.}
	\label{tab:G288732_obs}
	\scriptsize
	\begin{tabular}{ccclcccll}
		\hline
		mini-GWAC & $\mathrm{T_{start}}$ & $\mathrm{T_{end}}$  & ~~~~mid time & center RA & center dec & $\mathrm{P_{GW,cov}}$ & $\mathrm{N_{im}}$ / $\mathrm{N_{OT2}}$  & $\mathrm{M_{R,OT2}}$ \\
		grid / cam ID & 2017-06-08 & 2017-06-08  & ~~~~~ (hour)  & (h:m:s) & (deg:m:s) &  & & [min - max] \\
		\hline
		1 / C3 & 16:58:35 & 19:36:46 & $\mathrm{T_{GW}+16.2733}$ & 01:15:34.8 & +70:03:21 & 9.5$\%$ & 633 / 4& [9.9 - 8.8] \\
		2 / C4 & 17:10:44 & 19:31:01 & $\mathrm{T_{GW}+16.3267}$ & 01:22:21.1 & +49:56:18 & 0.5$\%$ & 561 / 0 & -- \\
		3 / C5 & 19:10:51 & 19:32:59 & $\mathrm{T_{GW}+17.3440}$ & 03:56:21.7 & +69:30:35 & 16.4$\%$ & 89 / 4 & [10.9 - 9.8] \\
		\hline
	\end{tabular}
\end{table*}

\end{document}